\documentstyle[12pt]{article}

\font\sevenrm=cmr7

\newcommand{\newsection}{
\setcounter{equation}{0}
\section}
\newcommand{\tr}{\,{\rm tr}\,}
\def\e{{\,\rm e}\,}

\def\eop{\vspace*{\fill}\pagebreak}
\def\be{\begin{equation}}
\def\ee{\end{equation}}
\def\bea{\begin{eqnarray}}
\def\eea{\end{eqnarray}}

\newtheorem{defin}{Definition}
\newtheorem{theorem}{Theorem}
\newtheorem{prop}{Proposition}
\newtheorem{prob}{Problem}
\newtheorem{claim}{Claim}
\newtheorem{corol}{Corollary}
\newtheorem{lemma}{Lemma}
\newtheorem{remark}{Remark}
\newtheorem{example}{Example}

\def\BQ{{\bf Q}}
\def\BZ{{\bf Z}}

\def\t#1{\widetilde{#1}}
\def\rom#1{{\rm{#1}}}
\def\G {\Gamma}
\def\g {\gamma}

\def\MM{{\widetilde M}}
\def\Bbb#1{{\bf {#1}}}
\def\ev#1{{\vec e}_{#1}}
\def\av#1{{\vec a}_{#1}}
\def\Tr{F}

\def\CUT{\hbox{\sevenrm cut}}
\def\Du{\Delta(u)}

\newcommand{\ie}{{\it i.e.,}\ }

\newcommand{\DD}{{\cal D}}

\newcommand{\supp}{{\hbox{supp\,}}}

\newcommand{\<}{\langle}
\renewcommand{\>}{\rangle}

\newcommand{\RR}{{\Bbb R}}
\newcommand{\HH}{{\Bbb H}}

\def\eps{\varepsilon}
\newcommand{\lvec}[1]{\stackrel{\leftarrow}{#1}}
\let\wtd=\widetilde
\let\pa=\partial

\title{{\bf \mbox{} \\Spectral problem on graphs\\
and $L$-functions}
\vspace{.5cm}} \author{{\bf L. O.
Chekhov}\thanks{E-mail: \ chekhov@mi.ras.ru} 
\\ \date{ } \vspace{.3cm} \\ {\it 
Steklov Mathematical Institute} \\ {\it Gubkin st.8, GSP-1, 117966 Moscow, 
Russia.}}

\begin{document}

\maketitle

\begin{abstract}
The scattering process on multiloop infinite $p+1$-valent graphs (generalized 
trees) is studied. These graphs are discrete spaces being quotients of the 
uniform tree over free acting discrete subgroups of the projective group 
$PGL(2, {\bf Q}_p)$. As the homogeneous spaces,
they are, in fact, identical to $p$-adic multiloop surfaces. 
The Ihara--Selberg $L$-function is associated with the finite 
subgraph---the reduced graph containing all loops of the generalized tree.
We study the spectral problem on these graphs, for which
we introduce the notion of spherical functions---eigenfunctions 
of a discrete Laplace operator acting on the graph. 
We define the $S$-matrix and prove its unitarity.
We present a proof of the Hashimoto--Bass theorem expressing $L$-function 
of any finite (reduced) graph via determinant of a local 
operator $\Delta(u)$ acting on this graph and relate the $S$-matrix 
determinant to this $L$-function thus obtaining the analogue of the 
Selberg trace formula. The discrete spectrum points are also determined 
and classified by the $L$-function.
Numerous examples of $L$-function calculations are presented.
\end{abstract}

\eop

\newsection{Introduction}
Two fundamental results relating the $L$-functions (zeta functions)
and Laplacian determinants have been obtained respectively
by Selberg~\cite{Sel}
(developed by Lax and Phillips~\cite{Lax}, 
L.~D.~Faddeev and B.~S.~Pavlov~\cite{FP}, A.~B.~Venkov~\cite{Venkov}, 
and others)
and Ihara~\cite{I}, Hashimoto~\cite{H-2}, and H.~Bass~\cite{B-2}.
The first result---the celebrated Selberg trace formula---concerns the
zeta functions and Laplacian determinants of compact (or noncompact with
finite area) in general multidimensional ($D$-dimensional)
manifolds of constant negative
curvature, which can be obtained from the $D$-dimensional upper half-space
with the hyperbolic metric
by factoring it over a discrete acting subgroup of the symmetry group 
of this space. We briefly expose these results in order to compare them
with the second class of results concerning the Ihara--Selberg
$L$-functions on finite graphs. 

The zeta function associated with a finite graph was introduced by
Ihara~\cite{I} who proved the structure theorem for torsion-free
discrete cocompact subgroups $\Gamma$ of the group $PGL(2,{\Bbb K}_p)$; \
${\Bbb K}_p$ being a $p$-adic number field or a field of power series 
over a finite field. Then, $\Gamma$ is a free group and the associated
zeta function can be constructed as follows. An element 
$\gamma\in\Gamma$, $\gamma\ne1$, is a {\it primitive} element if
it generates its centralizer in~$\Gamma$. Define the {\it length} 
$l(\gamma)\equiv\|\lambda_1/\lambda_2\|_p$ of the element where
$\|\cdot\|_p$ is the normalized valuation in ${\Bbb K}_p$ and
$\lambda_1,\lambda_2\in {\Bbb K}_p$ are the eigenvalues of~$\gamma$.
Let ${\cal P}(\Gamma)$ be the set of conjugate classes of primitive
elements of~$\Gamma$. Then, the Ihara zeta function is
\be
Z_\Gamma(u)=
\prod_{\gamma\in{\cal P}(\Gamma)}^{}
\bigl(1-u^{l(\gamma)}\bigr)^{-1}.
\label{&I1}
\ee
This definition was extended to finite-dimensional representations
of~$\Gamma$ in~\cite{I2}; then, instead of (\ref{&I1}), we have
\be
\label{&I2}
Z_\Gamma(u,\rho)=\prod_{\gamma\in{\cal P}(\Gamma)}^{}
\det\bigl(I_1-u^{l(\gamma)}\rho(\gamma)\bigr)^{-1},
\ee
where $\rho(\gamma)$ is the character of the given representation
and $I_1$ is the unit operator. (Original formula (\ref{&I1}) corresponds to
$\rho(\gamma)\equiv1$.)

Although zeta function (\ref{&I2}) is an infinite product, it was proved to be
a {\it rational} function. The Ihara theorem states that
\be
Z_\Gamma(u,\rho)=(1-u^2)^s\det(I_0-Au+pu^2),
\label{&I3}
\ee
where $s$ is an integer and $A$ is a matrix acting in a
finite-dimensional linear space. Using the rationality formula one can
counts the number of conjugate classes of primitive elements of~$\Gamma$.

{\bf 1.}
In~\cite{Ch1,Ch2} scattering processes on 
infinite graphs representing spaces of a constant negative 
curvature were studied. Such are uniform, or univalent
graphs, for which a natural number $p\ge 1$ exists
such that all vertices of the graph are incident exactly to
$p+1$ ends of edges. For prime~$p$, such graphs can be interpreted
as homogeneous spaces for the $p$-adic projective group 
$PGL(2,\BQ_p)$ factorized, first, by its maximal compact subgroup 
$PGL(2,\BZ_p)$ and, second, by some discrete, free-acting
(Schottky) group $\G$.
The homogeneous space $D\simeq PGL(2,\BQ_p)/PGL(2,\BZ_p)$ is a uniform
infinite tree graph. If we now factorize the tree $D$ by a discrete freely 
acting finitely generated  subgroup $\G_g\subset PGL(2,\BQ_p)$, 
where~$g$ is the number of the generating elements, then 
the obtained graph $T_g=D/\G_g$ is 
again a univalent graph  containing~$g$ loops and having
the tree~$D$ as the universal covering.

We present a proper analogue
of the Selberg trace formula \cite{Sel} for such discrete surfaces. These 
graphs sometimes are referred to as multiloop $p$-adic surfaces because
it has been shown in \cite{M-S}, \cite{CMZ} that for~$p$ the prime,
the discrete Laplacian action on these graphs yields the 
proper scattering amplitudes of the $p$-adic string. 

Note, however, that most of the calculations
were performed for trivial (Abelian) representation of the
group~$\Gamma$ in the $p$-adic string approach.

A natural distance $|P_{x,y}|$ over a path $P_{x,y}$ connecting
the points of a graph is
merely the number of edges entering (with multiplicities) this path.
One can consider linear spaces of functions $C_0$ and 
$C_1$ depending, respectively, on vertices $x_i$ and oriented 
edges $\ev j$ of the graph $T$.. A Laplace operator $\Delta$ acts on 
the space $C_0$ in a standard way,
$$
\Delta f(x)=\sum_{|P_{x,x_i}|=1}f(x_i)-(p+1)f(x).
$$
It is useful to segregate from the graph $T$ its 
finite ``closed'' part---the connected {\it reduced graph} $\Tr$ 
containing all internal loops. The valences of $\Tr$ 
vertices can be arbitrary ($\le p+1$). In~\cite{CMZ}, the string theory 
for such graphs was developed and the proper $p$-adic counterparts of
all crucial ingredients of the open string theory, such as prime forms, 
Schottky groups, etc. were found for the corresponding scattering amplitudes.

{\bf 2.} 
For the ordinary closed (or, more general,
finite area) Riemann surfaces of constant (negative) curvature
the Selberg trace formula,
which establishes an explicit relation between determinants of 
the Laplace operators and zeta functions (or, Ihara--Selberg $L$-functions),
is indeed close in form to (\ref{&I3}). The distance 
$l(\gamma)$ is now the length of a closed geodesics in the 
constant negative curvature metric. However, 
no proper analogue of the Selberg formula is known
for the essentially noncompact case where the spectrum of 
the Laplacian contains a continuous part responsible for the scattering
processes.
Note, however, that zeta function (\ref{&I1}) is 
still well defined for such noncompact surfaces; 
it is the Laplacian spectrum definition that hinders the progress in
this direction. One may hope that the graph description of
moduli spaces of complex curves with holes~\cite{Penner,Fock,Ch-Fock}
related to the geodesic description may help in finding analogous
statements for the rationality in the Riemann surface case.
We discuss this approach in Sec.~6.

{\bf 3.} We adopt the definition that zeta functions take values in the
field ${\Bbb k}_p$ (on the definition of the corresponding noncommutative
determinants, see~\cite{B-2}), while $L$-functions are assumed to take
values in the complex number field $\Bbb C$.

In a series 
of elegant papers by K.-I. Hashimoto \cite{H-2} and H.~Bass 
\cite{B-2} on zeta and $L$-functions,
the Ihara formula was generalized to the case of an arbitrary 
(not necessarily univalent) finite graph.
In the most general case, it has the form
\be 
L(u,\rho)=\det{}^{-1}(I_1-uT_1),
\label{&ration}
\ee
where the new operator $T_1$ is the operator acting on the space
of functions on oriented edges (simplexes of dimension one), which
corresponds to a translation along edges of the graph. 

Relation (\ref{&ration}) is suitable for both zeta and $L$-functions;
the difference is only in the definition of determinants (commutative
or noncommutative) involved. In this paper, we consider only the case
of $L$-functions, which seems to be physically meaningful as pertaining
to the spectral and scattering problems on graphs with potentials.

Formula (\ref{&ration}) is amazingly universal; it
expresses the {\it rationality condition},
which holds for the zeta functions in a host of cases
related to discrete dynamics---one may 
generalize it to the case of Hermitian (non-Abelian), 
Schr\"odinger type potentials on edges~\cite{Nov}, 
to the case of nonzero torsion
(potentials depending on ``angles'' along the path, i.e., on the pairs
of consecutive edges $(\vec e_i,\vec e_{i+1})$)~\cite{Kac-Sch}, 
etc. However, in most of these cases, we cannot find a bridge from
(\ref{&ration}) to the Laplacian determinants on nonuniform graphs. Only when
the potential $U_{\vec \mu}$
on (oriented) edges satisfies the unitarity condition,
i.e., $U_{\vec\mu}=U^{-1}_{\lvec \mu}$ for all pairs of oppositely
oriented edges, we can express the $L$-function in terms of Laplacian
type determinants on arbitrary (not necessarily univalent) graphs,
\be
L(u,\rho)=(1-u^2)^{s}\det{}^{-1}(1-uM_1+u^2Q) 
\equiv (1-u^2)^{s}\det{}^{-1}\Delta(u),\label{&0.2} 
\ee 
where $M_1$ is the above operator of summation over all neighbors 
(with probable potential dependence) and 
the new in comparison with (\ref{&I2}) element is the operator~$Q$,
which counts the number of these neighbors: $Qx=qx$ 
if the vertex~$x\in T$ has $q+1$ neighbors. In the trivial 
representation case, $s=|V|-|E|$, where
$|V|$ and $|E|$ are total number of vertices and 
(nonoriented) edges of the graph
(in a non-Abelian case, $s$ is the differences of
total (half)dimensions of linear spaces 
$v_{\cal A}^{2|V|}$ and $v_{\cal A}^{|E|}$, where elements of 
the representation $v_{\cal A}$ of the group or 
algebra ${\cal A}$ dwell on (now oriented) edges and vertices
of the graph.  However, the operator 
$\Delta(u)$ becomes the Laplacian only for $u=1$; these operators even do not 
commute at different~$u$. (For the detailed description of these results and 
their different applications, see~\cite{VN}.) 

{\bf 4.} The graphs~$T$ ($p$-adic multiloop surfaces) are noncompact in 
general, which differs them 
both from the closed Riemann surfaces and with finite graphs. 
However, it is possible to use the spherical function technique
and describe the problem in physical terms of the scattering
theory on graphs in this case.
In the original setting, spherical 
functions $\psi$ are eigenfunctions of the Laplace--Beltrami operator that 
depend only on the distance to a given point $x_0$ (the center). 
Because the 
Laplacian is the second-order operator, we always have two branches of the 
solution (at a distant point) proportional to $\alpha_+^d$ 
and $\alpha_-^d$ where~$d$ is the
distance to the center. 
Resolving the eigenvalue problem at the 
central point we fix the ratio of coefficients $a_+$ and $a_-$ 
standing by these two branches. If we choose $\psi(x_0)=1$, then $a_+$ 
and $a_-$ become Harish-Chandra coefficients and $c=a_+/a_-$ is a 
scattering amplitude of the $s$-wave.  Spherical functions have been 
found for the scattering on a quantum hyperplane \cite{VK} as well as 
for the scattering on $p$-adic hyperbolic plane \cite{Fr}. The 
$S$-matrices obtained are closely related to the partition functions 
of the $XXZ$ model and a number of nice but still mysterious relations 
between them have been obtained by Freund and 
Zabrodin \cite{FZ}.  

Investigating the spectral theory on graphs has already rich history.
Results obtained were related to studying eigenvalue problem
on finite uniform (\ie univalent) graphs. Here, it was proved that
a deep relationship exists between the modular forms on Teichm\"uller
spaces and finite graphs, namely, the so-called Ramanujan graphs. 
We do not discuss this interesting approach in this paper and instead
refer the reader to the monograph~\cite{Sarnak} and 
papers~\cite{Pizer,Erdos,VN}.

In~\cite{Ch1}, an analogue of a spherical function for the
multiloop graph was introduced. The main idea is the following. 
Considering the spectral problem $L\psi=\lambda\psi$, \ $L$ being the
Laplacian, we note that
any linear superposition of spherical functions with the same 
eigenvalue $\lambda$ but different scattering centers is again an 
eigenfunction. An eigenfunction of the Laplace 
operator on the factorized tree $T=X/\G_g$ corresponds to a source 
distribution function, $s(x)$, on the tree $X$ such that for every 
$\g\in\G_g$ and $x\in X$, we have $s(\g x)=s(x)$. Then the whole 
eigenfunction is periodic under the action of $\G_g$. Moreover, we 
choose a finite domain (a reduced graph) $F\subset T$ and consider only 
$s(x)$ such that $\supp s(x)\subseteq F$. 
Inside $T$ there is a unique minimal finite 
connected subgraph containing all loops---the union $D(\Gamma)$
of invariant axes
of all elements of $\Gamma_g$ factorized over the action of the group 
$\Gamma_g$.
This graph, $D(\Gamma)/\Gamma$,
 contains all information about the ``geometrical 
structure'' of $T$. We always assume $D(\Gamma)/\Gamma\subseteq F$.

Each eigenfunction $\psi(x)$ may be presented as a sum of 
retarded and advanced wave functions:
\be 
\psi(x)=A_{adv}(u)\alpha_+^{d(x,u(x))}-
A_{ret}(u)\alpha_-^{d(x,u(x))}\equiv \psi_+(x)-\psi_-(x),\label{&0.4}
\ee
where $\alpha_+$ and $\alpha_-$ are two fixed complex numbers 
depending only on the eigenvalue $t$ of the Laplacian $L\equiv\Delta(1)$, 
which acts on the whole graph~$T$, $L\psi=(t-p-1)\psi$ and on the initial
prime number $p$, \ $\alpha_\pm=\frac 
t{2p}\pm\sqrt{\frac{t^2}{4p^2}-\frac 1p}$, $\alpha_+\alpha_-=1/p$, and
$u(x)$ is the closest to $x$ point of the reduced graph~$F$
(for $x\in F$, $u(x)=x$).
As in the central symmetric case, $\psi(x)$ depends only on distance to 
$F$ on branches outside~$F$.

Now we can define a scattering matrix $S$ for such system. As a basis 
we choose functions $A_{adv}^i(u)$ and $A_{ret}^i(u)$  that are 
nonzero only for some ($i$th) point of $F$. Then, determining
the asymptotic vectors $\psi_i^\pm$ that behave as $\alpha_\pm^n$ when
going along $i$th branch and are zero otherwise, we can 
asymptotically expand $\psi$ in the sum of basis vectors
$\psi_{i_a}^+ -\sum_{\{j_b\}}^{}s_{i_a,j_b}\psi_{j}^-$, where now
$i_a,j_b$ are multiindices indicating components~$a$ of the representation
of $v_{\cal A}$ on the $i$th branch. Then, the matrix $s_{i_a,j_b}$
has the natural sense of the $S$-matrix.

The determinant of the matrix $S$ depends on the spectral parameter,
contains the information about spectrum, and, moreover, is
directly connected with the $L$-function of the graph $T$. (As for 
eigenfunctions themselves, the Lax--Phillips approach was developed
for their description~\cite{Roman}.)

In order to find the determinant of $S$ we impose a 
restriction:  $A_{adv}(u)/A_{ret}(u)=\hbox{const}$ for all points 
$u\in F$. Imposing this condition at 
all---both external and internal points of $F$, we fix 
an arbitrariness in the splitting of $\psi(x)$ 
into the advanced  and retarded waves.

Because the central object---the reduced graph~$F$---is finite,
a finite set of possible eigenvalues of~the $S$-matrix, \ie
such constants $c_i$ (the letter ``c'' 
originates from Harish-Chandra $c$-function \cite{HC}) exists.
Their product $C$ is therefore the determinant of the $S$-matrix. We also 
called it the total $C$-function. We present the proof of the 
basic formula establishing the
relationship between $C$ and determinants of a local operator 
$\tilde\Delta (u)$ acting only on the reduced graph~$F$,
\be
\det S(t)=\left(\frac{\alpha_+}{\alpha_-}\right)^{r_0}
\frac{\det\tilde\Delta(\alpha_-)}{\det\tilde\Delta(\alpha_+)},
\label{&0.3}
\ee
where $r_0$ is the total dimension of linear space of functions 
at vertices of the reduced graph~$F$, which take values in $v_{\cal A}$.

On the contrary, the operator $\tilde\Delta(\alpha_{\mp})$ in (\ref{&0.3}) is 
taken from (\ref{&0.2}), $\tilde\Delta(u)=1+\t{Q}u^2-u\t M_1$ 
and it is determined
completely in terms of the reduced graph itself, not of the whole graph;
the only remaining dependence of the 
``big'' graph $T$ is contained in arguments $\alpha_+$ and $\alpha_-$ 
of the function $\tilde\Delta(\alpha_\pm)$. Comparing (\ref{&0.3}) and 
(\ref{&0.2}) we obtain the relation between Harish-Chandra total 
$C$-function and the $L$-function of the $p$-adic curve:  
\be 
C=\left(\frac{\alpha_+}{\alpha_-}\right)^{r_0}
\left(\frac{1-\alpha_-^2}{1-\alpha_+^2}\right)^{r_0-r_1}
\,\frac{L(\alpha_+)}{L(\alpha_-)},\label{&0.5}
\ee
where $r_0-r_1$ is the difference of total dimensions of spaces of
functions determined on vertices and (nonoriented) edges;
it is equal $(g-1)\times |v_{\cal A}|$, where $g$
is the number of loops (the genus) of the graph and $|v_{\cal A}|$
the representation dimension.
We also prove using the geometrical setting of~\cite{Nov} that the
$S$-matrix is unitary in the scattering zone.

From another point of view, 
the idea to consider a proper {\it product\/} of scattering coefficients 
calculated at different $p$ is originated 
from the adelic ideology. Such  products 
in scattering processes was first proposed in \cite{Fr}, where 
the product of $C$-functions for scattering on $p$-adic 
hyperplanes taken over all primes $p$ appeared to be connected with the
$C$-function of scattering on genus one modular figure considered by 
L.~D.~Faddeev and B.~S.~Pavlov \cite{FP}. The very general formulas 
concerning the scattering on symmetrical spaces and 
gamma-function technique can be found in \cite{GK}.
One could hope to find a proper adelic {\it products\/} of $L$-functions 
appearing in our approach in order to compare them with the ones for 
Riemann surfaces. 

We also discuss the eigenvalue problem as regarding
to the discrete part of the Laplacian spectrum and establish two 
important relations concerning the $S$-matrix. The discrete part of
spectrum may contain apart of ``customary'' discrete
eigenvalues corresponding to exponentially decreasing in branches
eigenfunctions also the so-called exceptional eigenvalues
corresponding to eigenfunctions that vanish identically on {\it all}
branches. We first prove that poles of $L$-function correspond to
normal discrete eigenvalues iff these poles {\it do not cancel}
each other in (\ref{&0.5}), \ie if $\alpha_+$ and $\alpha_-$ are not 
simultaneously the poles of the $L$-function. On the contrary, as
soon as such situation takes place, \ie there are such poles 
$\alpha_+$ and $\alpha_-$
of the $L$-function that $\alpha_+\alpha_-=1/p$, the exceptional discrete
spectrum appears.

The paper is organized as follows: Section~1 contains definitions, 
the interpretation of $p$-adic multiloop curves as graphs
and the action of the Shottky group on the initial tree graph.
In Sec.~2, we describe
the automorphic functions and potentials on these graphs and introduce
operators acting on the generalized tree simplicial complexes.
In Sec.~3, we 
describe $L$-functions associated with such groups, or, 
equivalently, with the reduced graphs. The theorem by Hashimoto and Bass is 
formulated and a proof is presented. In Sec.~4, we consider the
spectral problem, 
introduce the spherical functions on multiloop graphs, define
the corresponding $S$-matrix and show that its determinant
can be expressed as a ratio of two $L$-functions.
We prove the unitarity of the $S$-matrix and describe how to find
discrete spectrum eigenvalues using the $L$-function technique.
Examples for genus~1 and~2 are presented in Sec.~5 together with a 
simple algorithm for calculating $L$-functions in lower genera based 
on the Hashimoto--Bass theorem.
Eventually, in Sec.~6, we present a construction of graphs for
describing the Teichm\"uller spaces of complex curves
in the Poincar\'e uniformization picture and set the problem
of finding a proper analogue of the Selberg trace formula
for open Riemann surfaces.

\newsection{Definitions}

\subsection{Graphs and trees}
Let $p$ be a natural number and $D$ a uniform tree graph 
of order $p+1$, $V(D)$ and $\vec L(D)$ the sets of its vertices and
(oriented) edges. In what follows, we always consider {\it
oriented\/} edges, i.e.,  a two-dimensional subspace corresponds to each 
(nonoriented) edge. 
For each two points $x,y$ of the tree, the {\it distance\/}
$d(x,y)$  is equal to the length of the unique way connecting
these two points. 

\begin{defin} \label{&def1}{\rm
Let $T$ be a graph with finite number of loops and branches
(tails), $V$ the set of its 
vertices, and $\vec L$ the set of (oriented) edges. The graph~$T$ is a coset
of its universal covering tree, $D$, over the action of freely acting
subgroup~$\Gamma$ of the group of motion of the tree, $T=D/\Gamma$.
We denote by $\vec e$ and $\lvec e$ two edges from $\vec L$
with the opposite orientations. The modulus $|\cdot|$ of a set is the
cardinality (perhaps, infinite) of this set.
}
\end{defin}

\begin{defin}\label{&def4}{\rm
Let an oriented {\em path} $P_{x,y}$
in a graph~$T$ (or~$D$) be a (unique) sequence
(finite or infinite) $(\ev1,\dots,\ev{m})$ 
of consecutive
($d_0\ev{i}=d_1\ev{i-1}$, $1<i\le m$, where $d_0$ and $d_1$ are
the respective operators that project an oriented edge to the vertex
it starts or terminates)
oriented edges without backtracking (i.e., 
$\ev{i}\ne{\lvec e}_{i-1}$ for $1<i\le m$)
starting at the vertex~$x=d_0\ev1$ and 
terminating at the vertex~$y=d_1\ev{m}$. 
The {\em path length} $|P_{x,y}|$ is the number of edges 
entering the path. In the tree~$D$, the length of the path~$P_{x,y}$ is 
always the {\em distance} between the vertices~$x$ and~$y$.

The path $P_{x,y}$ is {\it closed\/} if
$x=y$. The {\it proper closed path\/} is a closed path with
$\ev1\ne{\lvec e}_{m}$ (for a path of nonunit length). 
}
\end{defin}
We say that an edge $\vec e_\mu$ {\em follows} an edge $\vec e$ if 
$\vec e_\mu$ starts at the vertex at which~$\vec e$ terminates,
i.e., $d_0\vec e_\mu=d_1\vec e$.

\begin{remark}
{\rm The set of proper closed paths with the marked (starting=terminating)
points removed (by using the forgetting mapping) is in a one-to-one
correspondence with the conjugate classes of the homotopy group
$\pi_1(T)$.
}
\end{remark}

\subsection{$\Gamma$-action on trees $D$}
We recall some facts about the construction of the 
action of finitely generated freely acting group $\Gamma$ on the tree 
$D$.

The tree graph~$D$ possesses a rich group of isometries---the
transformations preserving the distance on the graph. Considering a
``rigid'' graph, i.e., a graph with the fixed ordering of edges in each
vertex, and imposing the condition of the ordering preservation under
the action of the symmetry group, 
we reduce this huge group to the group of rotations and translations
along the axes (infinite lines) of the tree. 
(It is possible to interpret the boundary of the tree $D$ as a
$p$-adic projective plane $P_1({\Bbb Q}_p)$~\cite{M-S}. 
Then the full group 
of motions of the tree $X$ is the projective group $PGL(2,{\Bbb Q}_p)$.)
A rotation element (an analogue of an
elliptic element of the projective group) is a tree 
rotation about a vertex $x_0\in D$ through the angle $2\pi n/(p+1)$, 
$1\le n\le p$. This is not however the case we consider in this paper.
Another type of transformations is provided by elements that have no
fixed points inside the tree (free-acting elements).

Without losing the generality, we assume that
the group $\Gamma$ is a discrete free-acting 
finitely generating subgroup of 
$PGL(2,{\Bbb Q}_p)$, that is,
no fixed points inside the tree exist for all nonunit elements of
this group.
(Thus, it is an analogue of a Fuchsian
group in ordinary hyperbolic geometry.)
We consider in what follows only finitely generating
groups~$\Gamma$ and assume that the group has~$g$ generating elements.
Each element $\gamma\in\Gamma$,
$\gamma\ne 1$, induces a translation of the tree as a whole
along the {\it invariant axis} $D(\gamma)\subset D$ of the
element. The invariant axis $D(\gamma)$
is a unique infinite oriented path $\ldots\ev{-1}\ev0\ev1\ev2\ldots$  
that maps to itself under the action of $\gamma$:  
$\gamma(\ev{i})=\ev{i+l}$ shifting in the positive direction w.r.t.\ its
orientation. 
Here $l\equiv l(\gamma)$ is the {\it element length},
$l(\gamma)=\hbox{inf}\{d(x,\gamma(x))|\ x\in V(D)\}$, where obviously
the minimum is reached on the set $x\in D(\gamma)$. Explicitly, an
element $\gamma$ defines a translation on the distance $l(\gamma)$
along the line $D(\gamma)$. 
(We assume $l(\gamma)<\infty$.)\footnote{Note that no analogue of
parabolic element exists in this geometrical setting.}

A centralizer $Z(\gamma)\subset\Gamma$ of the element $\gamma\in\Gamma$ 
is a cyclic group. If, besides, $\gamma$ is a generator of $Z(\gamma)$, then
$\gamma$ is called a {\it primitive element\/} of $\Gamma$. It is useful 
to define a von Mangolt function $\Lambda(\gamma)=l(\varpi)$, where 
$\gamma=\varpi^m$ and $\varpi$ is a primitive element of $\Gamma$.

Let us consider now the subtree $D(\Gamma)$, which is
the union of all invariant axes of the elements of 
$\Gamma$.  For an element $\gamma$ having the length 
$l(\gamma)$, the sequence $\ldots\ev{-1}\ev0\ev1\ev2\ldots$ has the structure
$\ldots(\ev1\ev2\dots\ev{l})(\ev1'\ev2'\dots\ev{l}')
(\ev1''\dots\ev{l}'')\ldots$, where $\ev{i}^{(n)}$ are copies of the edge
$\ev{i}\in T$. Moreover, the sequence $\ev1\ev2\dots\ev{l}$ must be the
proper closed path in~$T$ (with possible repetitions, \ie some of $\ev{i}$
may have the same preimages in $T$).

Let us consider the set of conjugate classes $\{\gamma\}$ of the group
$\Gamma$:
\be
\{\gamma\}:\left\{\bigcup\limits_{\gamma, \omega\in \Gamma} 
\omega\gamma\omega^{-1}\in\Gamma \right\}.
\ee
For each element $\beta\in\{\gamma\}$ and each vertex $x\in V(D)$, we have
$\beta x=\omega\gamma\omega^{-1}x=y$; hence,
$(\omega^{-1}y)=\gamma(\omega^{-1}x)$. Therefore, for each element
$\beta\in\{\gamma\}$, there exists such $\omega\in\Gamma$ that the
invariant axis of~$\beta$ is an {\it image\/} 
of the invariant axis of the generating element~$\gamma$ 
under the action of $\omega^{-1}$: \
$D(\beta)=\omega^{-1}D(\gamma)$. Thus, there is 
a one-to-one correspondence between the
conjugate classes $\{\gamma\}$ and a set of all proper closed paths in
the graph $T$.

We call {\it primitive\/} conjugate classes $\{\varpi\}$ such 
classes $\{\gamma\}$ that are generated by the primitive elements of 
$\Gamma$.  Then, the proper closed path corresponding to $\{\varpi\}$ is 
the proper closed path in $T$ that has no subperiods.

All graphs we consider are obtained by factoring a tree~$D$ over the
action of a free-acting symmetry group.

\subsection{Reduced graphs and branches.}

\begin{defin} \label{&def2}{\rm
Let the {\em reduced graph}~$F$
be a finite, necessarily connected
subgraph $F\subset T$ containing all loops of the graph~$T$.
Its universal covering, $D_F$, is a subtree in~$D$.
}
\end{defin}

For the generality sake, we do not demand the reduced graph to be
inambiguously determined by the graph~$T$. However, we always can
segregate the {\it minimum} reduced graph, which is the intersection of
all reduced graphs admitted by the given graph~$T$. This minimum reduced
graph exactly coincides with the union of all $D(\gamma)$, $\gamma\in\Gamma$,
factorized over the action of the group~$\Gamma$. This union is a subtree
$D(\Gamma)\subseteq D$, but neither~$D(\Gamma)$ nor~$F$ must be uniform 
graphs.

The subgraph $D(\Gamma)/\Gamma$ and, correspondingly, a reduced graph~$T$
are finite graphs containing exactly $g$ loops, where~$g$ 
(the genus) is the number
of generating elements of the group~$\Gamma$.

Because a subtree~$D(\Gamma)$ does not coincide in general with the whole 
tree~$D$, a quotient $\Gamma\backslash X$ would be
an infinite graph in contrast to graphs in paper~\cite{I}. 

The graph $T$ can be presented in the form
$T=F\cup B(T)$, where all $g$ loops of the graph~$T$ are contained
in the reduced graph~$F$. Meanwhile, the minimum
reduced graph contains no terminal
points, i.e., such points that are incident to only one edge in $\Tr$.
The complement to~$F$, 
$B(T)$, is a (finite or empty) set of {\em branches\/} growing
from the vertices of~$F$ in a way to make the total valence of
a vertex of $T$ to be $p+1$. We always assume that the number of
branches growing from a vertex of the reduced graph is $p-q$, where
$q+1$ is exactly the incident number of the vertex w.r.t.\ the edges
of the reduced graph.
An example of such factorized
tree $T$ for $p=3$ and $g=1$ is presented in Fig.~1.

\begin{remark}
{\rm
Note that a set of proper closed paths of the graph~$T$ has the
natural analogue in the continuous projective geometry case---it is
a set of closed geodesics on the (open) Riemann surfaces.
}
\end{remark}

For each point $y\in T$, we define its {\it distance to the
reduced graph\/} $d(y,F)$  by the formula $\inf_{x\in F}d(x,y)$, which we
sometimes abridge to~$d(y)$. This minimum
is between the point $y$ and a unique point $x\in F$, which we
call an {\it image\/} $t(y)\in F$ of the point $y$.  
Each branch $B$ can be therefore naturally projected into its summit 
(in other terms, root, image) $t(y)\in F$.

%\eop

\phantom{xxx}

%% Fig.~1. An example of a factorized tree for $p=3$ and $g=1$. %%

\begin{picture}(190,2)(-50,85)

\put(100,50){\oval(40,40)[b]}
\put(100,50){\oval(40,40)[t]}
\multiput(100,70)(-20,-20){2}{\line(-1,1){20}}
\multiput(100,70)(20,-20){2}{\line(1,1){20}}
\multiput(60,30)(20,-20){2}{\line(1,1){20}}
\multiput(140,30)(-20,-20){2}{\line(-1,1){20}}
\multiput(100,70)(20,-20){2}{\circle*{3}}
\multiput(80,50)(20,-20){2}{\circle*{3}}
\put(88,82){\line(0,1){8}}
\put(88,82){\line(1,1){8}}
\put(112,82){\line(0,1){8}}
\put(112,82){\line(-1,1){8}}
\put(88,10){\line(0,1){8}}
\put(96,10){\line(-1,1){8}}
\put(112,10){\line(0,1){8}}
\put(104,10){\line(1,1){8}}
\put(68,62){\line(-1,0){8}}
\put(68,62){\line(-1,-1){8}}
\put(68,38){\line(-1,0){8}}
\put(68,38){\line(-1,1){8}}
\put(132,62){\line(1,0){8}}
\put(132,62){\line(1,-1){8}}
\put(132,38){\line(1,0){8}}
\put(132,38){\line(1,1){8}}
\multiput(88,82)(24,0){2}{\circle*{2}}
\multiput(88,18)(24,0){2}{\circle*{2}}
\multiput(68,38)(0,24){2}{\circle*{2}}
\multiput(132,38)(0,24){2}{\circle*{2}}
\multiput(79,97)(6,0){8}{\circle*{2}}
\multiput(79,3)(6,0){8}{\circle*{2}}
\multiput(53,29)(0,6){8}{\circle*{2}}
\multiput(147,29)(0,6){8}{\circle*{2}}

%\put(20,65){\makebox{$\lambda_1$}}
%\put(35,50){\makebox{$\lambda_2$}}
%\put(35,30){\makebox{$\lambda_3$}}
%
\end{picture}

\vspace{4.5cm}

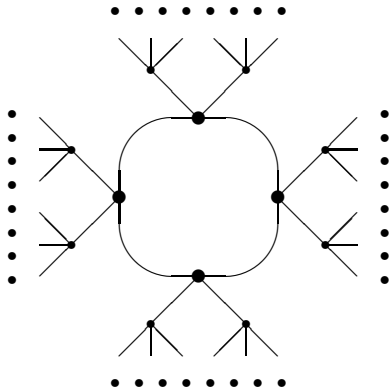
\centerline{{\bf Fig. 1.} An example of the factorized tree $T$ for
$g=1$, $p=3$.}

\vspace{6pt}

Note that the terminology of~\cite{Nov}, where the scattering theory on graphs
has been explored from the Schr\"odinger
potential standpoint, differs slightly from our
notation. The reduced graph is the {\it basis} graph
in Novikov's
notation. Moreover, instead of infinite branches growing from vertices
of~$F$, half-infinite tails growing from the corresponding points (nests)
have been considered. However, this setting is very close to
the one used here because we consider (automorphic)
functions that depend only on
distance to the reduced graph in what follows; in this respect, 
points of tails just label these distances and the both approaches are
equivalent in this respect.\footnote{An important difference is due to
{\em different} types of the potential operators considered.}

\newsection{Functions, operators, and potentials on graphs}

\subsection{Automorphic functions and potentials on $T$.}
In the original setting by Bass~\cite{B-2},
a unitary representation $\rho:\pi_1(T)\to GL(v_\rho)$ of the group $\Gamma$
was fixed and a linear space
$v_\rho$ was associated with any vertex (and, in principle, 
with any oriented
edge) of the tree~$D$. One can consider the space 
$L(\chi,D,v_\rho)$ of automorphic functions 
on the tree~$D$. Those are functions $G(x)$ defined on $V(D)$ (or
$\vec L(D)$), taking values in $V_\rho$, and satisfying the condition
$G(\gamma x)=\chi(\gamma)G(x)$ for all $\gamma\in \Gamma$ and $x\in D$.
Here $\chi(\gamma)$ is a left-invariant character of the group 
$\Gamma$. If $Y\subset D$ is a fundamental domain of $\Gamma$ in $D$ then 
each $G\in L(\chi,D,v_\rho)$ is completely defined by its values on $Y$.

However, we adopt an equivalent description, which is closer to
physically meaningful lattice gauge theories. Namely, instead of
considering automorphic functions, we consider 
periodic functions
on $D(F)$, which can be lifted to the
graph~$T$ (the factorized tree), while nontrivial character set of the
group~$\Gamma$ is ensured by introducing (unitary)
{\it potentials} on the edges of the graph~$T$. 

Let $\cal A$ be a group (or an algebra)
and $v_\rho$ the finite-dimensional representation
of $\cal A$.

In~\cite{Ch1}, the case of the trivial representation
$\chi(\gamma)\equiv 1$ was considered; nevertheless, already this
simplest example manifests all the main features
of the general theory. 
In the present paper, we demonstrate how the analogous method can be
applied to the case of arbitrary unitary (in general, non-Abelian) 
representation $(v_\rho,\chi)$.

\begin{defin}\label{&def3.1}{\rm
Given the group~$\cal A$ and its finite-dimensional representation~$v_\rho$,
we associate the representation space~$v$ of the group ${\cal A}$ to each
vertex and to each oriented edge 
of the graph~$T$. Then, the corresponding linear spaces
of {\em functions} defined on vertices or edges of a graph
and taking values
in~$v_\rho$ are $C_0=v^{|V|}$ and $C_1=v^{|\vec E|}$
(the power is understood as the tensor product power).

In what follows, we interpret tilde quantities as pertaining to 
the reduced graph or to its universal covering. For example,
the linear spaces $\wtd C_0$ and $\wtd C_1$ correspond to vertices
and edges of the reduced graph~$F$. Note that the graph is always
assumed to be {\em closed}
as the simplicial complex, \ie every edge enters this graph
together with its both endpoints (vertices).

We define also the natural bilinear forms on $C_0$ and $C_1$: 
for $f,g\in C_0$,
\be
\<f,g\>=\sum_{x\in V}^{}f(x)^\ast g(x),\quad f(x),g(x)\in v_\rho,
\label{&brackets}
\ee
and analogously for $f,g\in C_1$.
} 
\end{defin}

\begin{defin}\label{&def3}{\rm
{\bf 1.} First type of potentials are the {\em unitary
potential on edges}, $U_{\vec e}\in {\cal A}$, where $\cal A$ 
is a (non-Abelian) group, the unitarity condition implies that
$U_{\vec e}=U^{-1}_{\lvec e}$. Those are the potentials one
encounters in the Bass case~\cite{B-2}.

{\bf 2.} Second type of potential are
{\em Hermitian potentials $A_{\vec e}$
on edges} lying in an {\it algebra} $\cal A$ such that
$A_{\vec e}=A_{\lvec e}=A^+_{\vec e}$ (the reality condition)~\cite{Nov}.

{\bf 3.} Third, we call the {\em nontrivial torsion} potentials the 
potentials $U_{\vec\mu\vec\nu}$ determined for the {\em pairs} of 
consecutive edges ($\vec \mu$ precedes $\vec\nu$) such that
\be
U^{-1}_{\vec\mu\vec\nu}=U_{\lvec\nu\lvec\mu}.
\label{&**.1}
\ee
}
\end{defin}

Another natural demand, which is often imposed on the set of torsion
matrices $U_{\vec\mu\vec\nu}$ is as follows. Let $\mu_i$, $i=1,\dots,n$,
be edges coming (in the fixed cyclic ordering) into a vertex of the graph.
Then, for all $1\le i,j,k\le n$,
\be
U_{\vec\mu_i\lvec \mu_k}U_{\vec\mu_k\lvec \mu_j}=U_{\vec\mu_i\lvec \mu_j};
\label{&**.2}
\ee
these relations must hold for any vertex and they imply, in particular, that
$U_{\vec e_i\lvec e_i}=I_1$ for any edge $\vec e\in\vec L(D)$.

\subsection{Hecke operators on graphs}

In this section, $C_0$ and $C_1$ are the corresponding spaces of functions
for {\em arbitrary} graph.

We consider first the operators that act inside linear
spaces $C_0$ and $C_1$. All these operators are assumed to act on the
corresponding universal covering~$D$ (or $D_F$)
with the natural identification of (oriented) edges
and vertices of~$D$ with their preimages in~$T$ (or in $F$). We denote by
$I_0$ and $I_1$ the identity operators in the respective 
spaces $C_0$ and $C_1$.

{\bf1.} We introduce the set of the Hecke
operators $M_n: C_0\to C_0$, $n=1,\dots, \infty$, such that
their action on basis vectors is
\be
M_nv_x=\sum_{y:\,|P_{x,y}|=n}^{}U_{\vec\mu_1}\cdots U_{\vec\mu_n}v_y,
\label{&Mn}
\ee
where the product runs over all oriented edges entering the path $P_{x,y}$
and $x,y\in D$.

For the tilde operators pertaining to simplicial complexes
associated with the reduced graph~$F$, the corresponding to
(\ref{&Mn}) definition is
\be
\wtd M_nv_x \equiv\sum_{{y:\,|P_{x,y}|=n
\atop P_{x,y}\subset D_F}}^{}U_{\vec\mu_1}\cdots U_{\vec\mu_n}v_y.
\label{&Mn1}
\ee

The action of operators (\ref{&Mn}) and (\ref{&Mn1}) (as well as 
of all other operators) must be continued to the whole space $C_0$
by the linearity property.

{\bf2.} Next, we have the valency-counting operator~$Q:\, C_0\to C_0$,
\be
Qv_x=(\# \hbox{neighbors}-1)v_x.
\label{&Q}
\ee
For the uniform tree, the operator~$Q$ is merely the identity times~$p$.
However, for a nonuniform tree, say, for $D_F$, the operator~$Q$ possesses
some nontrivial properties.

{\bf3.} 
The $\Delta$-operators $\Delta(u): C_0\to C_0$ and 
$\tilde\Delta(u): \wtd C_0\to \wtd C_0$,
\be
\Delta(u)=I_0-uM_1+u^2Q, \qquad \tilde\Delta(u)=I_0-u\wtd M_1+u^2Q,
\quad u\in{\bf C},
\label{&D}
\ee
which have inverse operators (for $|u|<1/p$ in the trivial character case),
play an important role in the Bass construction.
We sometimes label their representation dependence writing them as, say, 
$\Delta_\rho(u)$.

Operators $M_n$ (\ref{&Mn}) 
constitute a basis in the center of endomorphism algebra.
Their multiplication algebra in the tree~$D$ is
\bea
M_1M_n&=&M_{n+1}+pM_{n-1},\quad n\ge2\nonumber\\
M_1M_1&=&M_2+(p+1)M_0,\quad M_0\equiv I_0.\label{&2}
\eea
If $\psi$ is an eigenvector of $M_1$ with the eigenvalue $t$,
\be
M_1\psi=t\psi.
\ee
then it is also an eigenvector of all $M_n$ and
\be
M_n\psi=S_n(p,t)\psi\label{&3}
\ee
where $S_n(p,t)$ is a system of orthogonal polynomials in $t$ with a
generating relation $tS_n(p,t)=S_{n+1}(p,t)+pS_{n-1}(p,t)$, $S_1=t$,
$S_2=t^2+p+1$.

In the general case of a nonuniform tree $D_F$, 
the operator~$Q$ obviously does not
commute with the Hecke operators, which in turn also become mutually
noncommutative. Nevertheless, 
it is possible to write algebraic relations 
analogous to (\ref{&2}) using the operator $Q$:
\bea
\MM_1\MM_n&=&\MM_{n+1}+Q\MM_{n-1},\quad n\ge2\nonumber\\
\MM_1\MM_1&=&\MM_2+(Q+1)\MM_0,\quad \MM_0=\hbox{id}.\label{&MQ}
\eea
Moreover,
\be
\sum_{n=1}^{\infty}u^n\wtd M_n \tilde\Delta(u)=(1-u^2)I_0.
\label{&a1}
\ee
Relation (\ref{&a1}) implies that the operator $\tilde\Delta(u)$ is the 
{\it generating function} for the operators $\wtd M_n$.

{\bf4.} The {\it Laplacian} of a unitary theory is merely $\Delta(1)$.
For this Laplacian to be a Hermitian operator w.r.t.\ brackets
(\ref{&brackets}) we must impose the unitarity condition on the
potential,
\be
U^+_{\vec \mu}=U^{-1}_{\vec \mu}.
\label{&unit}
\ee

In~\cite{Nov}, the Laplacian was determined as the operator of the
Schr\"odinger type, \ie it has form (\ref{&D}) with the first Hecke
operator~$M_1$ as in (\ref{&Mn}) but with the {\it Hermitian} potential
$A_{\vec\mu}\equiv A_{\lvec\mu}$ (elements of an algebra)
substituted for $U_{\vec\mu}$ (elements of a group)
on the edges of a graph.
The corresponding Laplacian 
\be
\Delta v_x=\sum_{y:|P_{x,y}|=1}^{}\bigl(A_{\vec\mu_{xy}}v_y-v_x \bigr)
\label{&Lap}
\ee
is also Hermitian, but no relations of
type (\ref{&MQ}) exist.

{\bf5.}
On the space $C_1$ we first define the inversion map 
$J:\,C_1\to C_1$, which merely
changes all orientations of edges,
\be
J{\vec e}=U_{\vec e}{\lvec e},
\label{&J}
\ee
Moreover, there exist the set of Hecke operators for the space $C_1$. 
They all are generated by a single operator $T\equiv T_1:\,C_1\to C_1$.

\begin{defin}\label{&TT}
The first Hecke operator acting on the space $C_1$ is
\be
T{\vec e}=\sum_{\vec e_\mu\hbox{\,\scriptsize
following\,}\vec e}^{}A_{\vec e}\,{\vec e}_\mu.
\label{&T}
\ee
\end{defin}

Then, we can obviously 
define $T_n$ as the sum over all terminal edges of 
oriented reduced paths of the length $n+1$ starting from the given edge,
\be
T^m(\ev0)=\sum_{(\ev0,\ev1,\dots,\ev{m})_{red}}^{}\,\ev{m}.\label{&Tm}
\ee
However, in contrast to the space
$C_0$, the relation between $T_n$ and $T_1$ is merely $T_n=T_1^n$ 
for {\em any} graph, and
the family of these operators is commutative for every tree, 
no matter uniform or nonuniform. Then, for any graph,
the corresponding 
generating function is $I_1-uT_1$:
\be
\sum_{n=0}^{\infty}u^nT_n(I_1-uT_1)=I_1,\quad u\in{\bf C}.
\ee

Definition~\ref{&TT} can be easily generalized to other potentials from
Definition~\ref{&def3}. For $U_{\vec\mu}$ replaced by the Hermitian matrices
$A_{\vec\mu}$, we just obtain the Novikov~\cite{Nov} potential on edges.
Important examples are provided by potentials of the third type (with the
nontrivial torsion). Then, in the most general case, we can define~$T$ as
\be
T\vec e=\sum_{\vec e_\mu \hbox{\sevenrm\ following\ }\vec e}^{}
U_{\vec e}F_{\vec e\vec e_\mu}\vec e_\mu,
\label{&***.1}
\ee
where neither $U_{\vec e}$ nor $F_{\vec e\vec e_\mu}$ must satisfy the
unitarity or Hermiticity conditions; in formula (\ref{&***.1}), the 
potentials on edges with opposite orientations can be set {\em arbitrary},
the $L$-function expression through the operator~$T$, which are
presented in the following section, remain nevertheless valid.

We note two important examples of (\ref{&***.1}). 
The graph description~\cite{Penner,Fock} of the
Teichm\"uller spaces of complex Riemann surfaces discussed in Sec.~6;
the other is the
Ising model on a lattice governed by the Kac--Ward operator 
determinant~\cite{Kac-Sch}.

\subsection{Intertwining operators between $C_0$ and $C_1$.}

Following \cite{B-2}, we introduce a set of operators acting
between two spaces $C_0\leftrightarrow C_1$. 

{\bf1.} First, we have two {\it boundary
mappings\/} $\pa_1,\pa_0:\,C_1\to C_0$,
\be
\pa_1{\vec e}=A_{\vec e}x_1
\label{&p_1}
\ee
and
\be
\pa_0{\vec e}=x_0,
\label{&p0}
\ee

{\bf2.} We also have
the coboundary operator $\sigma_0:C_0\to C_1$,
\be
\sigma_0x=\sum_{|P_{x,y}|=1}^{}{\vec e}_{(x,y)},
\label{&s0}
\ee
which set into the correspondence with the point~$x$ a linear
combination of {\it all} edges outgoing from this vertex.
(In~\cite{B-2}, the operator $\sigma_1$, which sets
into the correspondence to the vertex~$x$ all {\it incoming\/} edges
was defined.
However, this definition, which is meaningful for a nonpotential case,
becomes ill defined when a potential is introduced. Fortunately,
in what follows, we do not need the operator $\sigma_1$.)

\subsection{Relations between operators}

Let us introduce the total (half-)dimensions of the spaces $C_0$ and $C_1$,
\be
r_0=\hbox{rank\,}(C_0),\qquad r_1=\hbox{rank\,}(C_1)/2.
\ee
We also introduce auxiliary operators
\be
\pa(u)=\pa_0u-\pa_1;\qquad \sigma(u)=\sigma_0 u.
\ee
The standard boundary operator is then $\pa\equiv\pa_0-\pa_1=\pa(1)$.

Direct calculations show that the above
operators satisfy the following
relations~\cite{B-2}, which hold for any graph:
\bea
&{}&\pa_0\sigma_0=Q+1;\nonumber\\
&{}&\pa_1\sigma_0=M_1;\nonumber\\
&{}&T=\sigma_0\pa_1-J;\label{&formulas}\\
&{}&\pa(u)\sigma(u)=\Delta(u)-(1-u^2)I_0;\nonumber\\
&{}&\sigma(u)\pa(u)=u(T+J)(uJ-I_1).\nonumber
\eea

An important assertion proved by Bass establish a connection
between the spectral Laplacian problem and the spectral problem
for the Hecke operator~$T_1$.

\begin{lemma}\label{&lem1}~{\rm\cite{B-2}}
For the potentials of the first type in Definition~{\rm\ref{&def3}},
we obtain
$$
\det\limits_{C_1}(I_1-uT_1)=(1-u^2)^{r_0-r_1}\det\limits_{C_0}\Delta(u).
$$
\end{lemma}
{\em Proof}. Let us define the matrices acting in $C_0\oplus C_1$,
$$
L=\left[\begin{array}{cc}
       (1-u^2)I_0&\pa(u)\\
          0      & I_1\end{array}\right],\qquad
M=\left[\begin{array}{cc}
        I_0&-\pa(u)\\
  \sigma(u)& (1-u^2)I_1\end{array}\right].
$$
Then from (\ref{&formulas}), we obtain
$$
LM=\left[\begin{array}{cc}
        \Delta(u)&0\\
  \sigma(u)& (1-u^2)I_1\end{array}\right]\hbox{\ and\ }
ML=\left[\begin{array}{cc}
        (1-u^2)I_0&0\\
  \sigma(u)(1-u^2)& (I_1-uT)(I_1-uJ)\end{array}\right].
$$
The assertion of Lemma \ref{&lem1} follows from equating the traces
of $LM$ and $ML$; we only need to evaluate
the determinant of $I_1-uJ$. It is block--diagonal in the basis of 
(nonoriented) edges. Each edge admits two orientations,
so we have
\be
\det (I_1-uJ)=\det \left[\matrix{I_1 & -uU\cr -uU^{-1} & I_1}
\right]^{|\vec E/2|}=(1-u^2)^{r_1}_{}
\ee
for any matrix~$U$.
Note that $r_1$ is exactly the dimension of the representation
of the group~$U$ times $|E|$, where~$|E|$ is the number of unoriented
edges. Therefore, $r_0-r_1=\dim U\times (|V|-|E|)$ where the difference
between the numbers of vertices and (nonoriented) edges is 
exactly~$1-g$---the Euler characteristic of the graph.

In general, no relation of type of Lemma \ref{&lem1} exists for potentials
of the second and third type from Definition \ref{&def3}. Therefore, the 
spectral problems for the operators $\Delta(u)$ and $T(u)$ are not
related in these cases. However, for graphs of the third kind from
Definition~\ref{&def3} with conditions (\ref{&**.1}) and (\ref{&**.2})
imposed, we can, nevertheless, connect these two operators for the price
of ``blowing up'' the vertices of a graph; this construction
is discussed in Sec.~6.

\newsection{$L$-function}

\begin{defin}{\rm~\cite{B-2}
Let $v_\rho$ be a $\bf C^k$-module with the character
$\chi_\rho:T(C[T])\to \bf C^k$. Then, 
the {\em Ihara--Selberg $L$-function} $L(\rho,u)$ is
\be
L(\rho,u)=\prod_{\{\varpi\}\in\Gamma}^{}\det\bigl(I_v-u^{l(\varpi)}
\rho(\varpi)\bigr)^{-1},
\label{&Lff}
\ee
where the product ranges all conjugate classes of primitive
elements of the group~$\Gamma$, or, equivalently, all proper
oriented closed paths without periods in the graph~$T$,
$l_\eps$ are lengths of these paths, and 
$\rho(\varpi)=U_{\vec e_1}\dots 
U_{\vec e_{l_\eps}}$ is the product over the path that corresponds
to the element $\varpi\in\Gamma$ of elements of the group $\cal A$.
}
\end{defin}

We present now the Bass' formulation and proof~\cite{B-2} of the 
Hashimoto's theorem~\cite{H-2}.

\begin{theorem}\label{&th1}
Let $\rho:\Gamma\to GL(V)$ be a finite-dimensional complex representation 
of~$\Gamma$. Then, the $L$-function {\rm(\ref{&Lff})} is a rational function
in the variable~$u${\rm:}
\be
L(\rho, u)^{-1}=\det\limits_{C_1}(I_1-uT_1),
\label{&L.v.T.}
\ee
where $T_1$ is from {\rm(\ref{&T})}.
\end{theorem}

From Lemma~\ref{&lem1} and Theorem~\ref{&th1} the corollary follows

\begin{corol}\label{&cor1}
If the unitarity condition {\rm(\ref{&unit})} is satisfied,
then the $L$-function can be expressed through the determinant
of the operator $\Delta_\rho(u)${\rm:}
\be
L(\rho, u)^{-1}=(1-u^2)^{r_0-r_1}\det\limits_{C_0}\bigl(\Delta_\rho(u)\bigr).
\label{&t1}
\ee
\end{corol}

The product in (\ref{&Lff}) converges absolutely in the domain $\{u:\,
|u|<(q_{max}|{\cal A}|)^{-1}\}$, 
where $q_{max}\le p$ is the maximum incident number of the
tree $D_F$ and $|{\cal A}|$ is an absolute value of a maximum element
$U_{ij}^{\mu}$. 
Also, we have
for the logarithmic derivative of $L(\rho, u)$,
\be
\frac d{du}\log \bigl(
L(\rho,u)\bigr)=\sum_{\{\gamma\}}^{}\Lambda(\gamma)
\tr (\rho(\gamma))u^{l(\gamma)-1},\label{&1.2}
\ee
where the sum ranges all conjugate classes $\{\gamma\}$ distinct
from $\{1\}$.

{\bf Proof of Theorem~1.}

Let us choose in the tree $D$ some {\it fundamental domain\/}
$\DD(\Gamma)$ of the symmetry group $\Gamma$. We now consider the
contribution to the trace of the operator $T^m$ coming
from some edge $\ev1\in
\DD(\Gamma)$. Only those $\ev{i}\in D$ contribute to $\tr T^m$
that
\begin{enumerate}
\item lie on the distance $m$ to the initial edge $\ev1$ (along some reduced
path $\ev1\ev2\dots\ev{m+1}$);
\item have the orientation along this path.
\end{enumerate}
The crucial observation follows. Let us consider the action of some element
$\gamma\in\Gamma$ on oriented edges of $D$ (Fig.~2). It is easy to see that
if the edge does not belong to the invariant axis $D(\gamma)$ of this
element (like $\ev{x}$ does not on Fig.~2), then the orientation of its image 
under the action of the element~$\gamma$,
$\gamma\ev{x}$, is {\it opposite\/} to the orientation of the reduced 
path coming from $\ev{x}$ to $\gamma\ev{x}$. Therefore, no
such edges contribute to the trace! 
Only the edges that (like $\ev1$ in Fig.~2) belong to the
invariant axis $D(\gamma)$ preserve their orientations 
toward $D(\gamma)$ under the action of~$\gamma$ (the translation along
$D(\gamma)$). Say, for $\ev1$, we have $\ev{l+1}=\gamma\ev1$ in Fig.~2.

\eop

\phantom{xxx}
\begin{picture}(190,2)(0,85)

\multiput(20,20)(25,0){7}{\vector(1,0){25}}
\multiput(45,20)(125,0){2}{\circle*{2}}
\put(195,20){\line(1,0){50}}
\multiput(45,20)(125,0){2}{\line(-1,2){30}}
\multiput(45,20)(125,0){2}{\line(1,2){30}}
\multiput(15,80)(125,0){2}{\circle*{2}}
%\multiput(25,60)(125,0){2}{\circle*{2}}
%\put(160,40){\circle*{2}}
\multiput(15,80)(125,0){2}{\vector(1,-2){10}}
\multiput(170,20)(-10,20){2}{\vector(-1,2){10}}
\put(50,10){\makebox{$\ev1$}}
\put(175,10){\makebox{$\ev{l+1}\!=\!\gamma\ev1$}}
\put(230,10){\makebox{$D(\gamma)$}}
\put(20,75){\makebox{$\ev{x}$}}
\put(145,75){\makebox{$\gamma\ev{x}$}}
\put(155,55){\makebox{$T^m\ev1$}}

%\put(20,65){\makebox{$\lambda_1$}}
%\put(35,50){\makebox{$\lambda_2$}}
%\put(35,30){\makebox{$\lambda_3$}}
%
\end{picture}

\vspace{4.5cm}

\centerline{{\bf Fig. 2.}
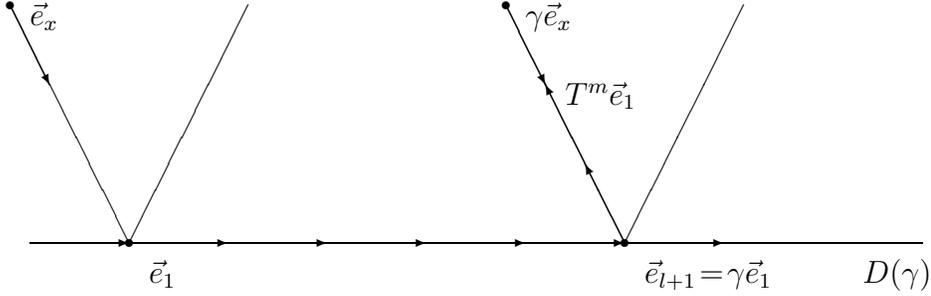
The action of the group element $\gamma\in\Gamma$ on the
edges of $D$,}
\centerline{with the length $l(\gamma)=l$. Note that $\gamma\ev{x}\ne
T^{m+1}\ev1$ for all $m$.}

\vspace{5pt}

{\it Note\/}. This is why we use $\tr T^m$ instead of, say, $M_m$---the 
$m$th Hecke operator for the space $C_0$---for $\tr M_m$, 
{\em all} pairs of vertices $x,y\in D$, not only
those belonging to the invariant axis $D(\gamma)$,
contribute to this trace as soon as $d(x,y)=m$ and $y=\gamma x$.

Coming back to the space $C_1$, we fix for a moment the edge
$\ev1\in D(\Gamma)\subseteq D$.
As was mentioned above, for each
primitive conjugate class $\varpi$ one can find
not necessarily one axis
$D(\varpi)\subset D_F$ such that $\ev1\in D(\varpi)$. We denote by
$\av{i}$  the oriented edges of the reduced graph $\Tr$ itself, thus each
$\ev{i}\in D_F$ is an {\em image} of the edge $\av{i}\in \Tr$ with the
orientation naturally preserved. 
For each element $\gamma\in \Gamma$, we then
set into the correspondence the periodic
sequence of ``letters'' (the cyclic word)
\be 
D(\gamma)\simeq 
\ldots(\av1\av2\dots\av{l})(\av1 \av2\dots\av{l})\ldots, 
\label{&D(sigma)} 
\ee 
where $l\equiv\Lambda(\gamma)=l(\varpi)$ is the
length of the generating element for $\gamma$. 

Note that the elements of $\Gamma$ determine the
symbolic dynamics of ``words'' ({\em Lyndon words}, see~\cite{Foata})
composed from the ``letters''
$\av{i}$, which can be noncommutative elements of the group ${\cal A}$.
We even do not need a unitarity condition (\ref{&unit}) in this case;
in particular, Theorem \ref{&th1} holds even if 
letters $\av{i}$ and $\lvec a_i$ are not related. Thus, for the rest of
the proof, we
merely identify the letters $\av{i}$ with the elements $U_{\ev{i}}\ev{i}$
and the proof is valid for any (non-Abelian) group or algebra
elements dwelling
on the oriented edges of the graph~$F$.

It is clear that if
$l(\gamma)=m$, then $l(\varpi)=l$, $l|m$, that is, $l$ is a divisor of $m$.
Note again that there can be repeated $\av{i}$ in the cyclic word
$\av1\av2\dots\av{l}(\av{l+1}\equiv\av1)$,
Moreover, this sequence may contain subperiods if $\gamma$ is not a
primitive element, but, obviously, the minimum length of this subperiod, $l$,
is a divisor, $l|m$.

We now fix the element $\av1$---the preimage of $\ev1$. In order to determine
its contribution to $\tr T^m$ we must find all possible {\em different} 
cyclic expressions
\be
\av1\av2\dots\av{m}(\av{m+1}=\av1) \label{&av1--avm}
\ee
that include this element.
We now establish a correspondence between sets (\ref{&D(sigma)}) and
(\ref{&av1--avm}).

{\bf 1.} To each finite cyclic sequence $\av1\dots\av{m}(\av1)$ we 
inambiguously set into the correspondence
the infinite periodic sequence $\dots(\av1\dots\av{m})
(\av1\dots\av{m})\dots$ corresponding to a unique element of primitive
conjugate class $\{\varpi\}$ with $l(\varpi)=l|m$.

{\bf 2.} On the contrary, we now
choose an element from $\{\varpi\}$, or,
equivalently, from some periodic reduced sequence
\be
\ldots(\dots)(\av{i_1}\av{i_2}\dots\av{i_l})(\av{i_1}\dots)\dots
\label{&p1}
\ee
with no subperiods and with the minimal period $l=l(\varpi)$. If it contains
the (oriented) edge $\av1$ \ $d_1$ times among the edges
$\{\av{i_1},\dots,\av{i_l}\}$ and, moreover, $l$ is the divisor of $m$, then
there are exactly $d_1$ {\it different\/} sequences $\av1\dots\av{m}(\av1)$
(\ref{&av1--avm})
containing in (\ref{&p1}). Eventually, doing the sum over all edges of $\Tr$
(or of the fundamental domain $\DD(\Gamma)\in D$), \ie evaluating the trace 
of the operator $T^m$, we find that the contribution from the sequence 
$\ldots(\dots)(\av{i_1}\dots\av{i_l})\dots$ to this trace is 
\be
\sum_{j=1}^{|\vec L_F|}\#\left\{\av{j}\ \hbox{in}\
\{\av{i_1},\dots,\av{i_l}\}\right\}\equiv l,
\ee
where
$l$ is the total length of the primitive element $\varpi$. Therefore, we
have the following remarkable formula:
\be
\tr T^m=\sum_{l|m}^{}l\cdot 
\sum_{\{\varpi:l(\varpi)=l\}}^{}\tr\bigl[\rho(\varpi)^{m/l}\bigr],
\label{&T1m}
\ee
where the sum runs over all primitive conjugate classes of $\Gamma$.

We now obtain from (\ref{&1.2}) the representation for the $L$-function:
\bea
u\frac d{du}\log \bigl(
L(\rho,u)\bigr)&=&
\sum_{\{\varpi\}}^{}\sum_{n=1}^{\infty}l(\varpi)\cdot
u^{nl(\varpi)}\tr\bigl[\rho(\varpi)^n\bigr]\nonumber\\
&=&\hbox{from\ (\ref{&T1m})}\ \sum_{k=1}^{\infty}\tr T^k\cdot u^k
=-u\frac{d}{du}\log\,\det(1-u\cdot T).\label{&LT}
\eea
This completes the proof of the theorem.

\newsection{Spectral problem on infinite graphs}
We set the spectral problem considering a function $\psi\in C_0$
such that
\be
L\psi=\lambda\psi,\quad \lambda\in{\bf C},
\label{&*}
\ee
where the Laplacian $L$ is
\be
L=M_1-(Q+1)..
\ee
We first determine the solution of (\ref{&*}) on a branch (tail). 
We assume 
the graph~$T$ to be uniform ($(p+1)$-valent) but infinite. Then, we consider 
a branch (tail) of the graph~$T$ with the starting vertex~$x_0$ (the 
index~$i$ is the distance to~$x_0$ in the reasoning below). We have
$\psi(x_i^{(s)})=v_i^{(s)}$; then,
\be
\sum_{s=1}^{p}v_{i+1}^{(s)}+v_{i-1}^{(\cdot)}-(p+1)v_i=\lambda v_i,\quad i>0,
\label{&*2}
\ee
where $x_{i-1}^{(\cdot)}$ is a unique vertex (from the $(i-1)$th layer)
that precedes~$x_i$ (when moving along the branch from its root).

The general solution to (\ref{&*2}) is
\be
v_i^{(s)}=\sum_{\pm}^{}\alpha_\pm^i \bigl(A_{i-1}^{(s_{i-1})}\bigr)^{-1}
\dots \bigl(A_{1}^{(s_1)}\bigr)^{-1}v_\pm^0,
\label{&*3}
\ee
where the product ranges all edges of the path $P_{x_0,x_i^{(s_i)}}$,
$v_+^0$ and $v_-^0$ are constant vectors from~$V$, and $\alpha_\pm$ are
the solutions of the equation
\be
p\alpha_\pm^2-(p+1+\lambda)\alpha_\pm+1=0;\quad 
t\equiv \frac{p+1+\lambda}{2p},\quad \alpha_\pm=t\pm \sqrt{t^2-1/p}.
\label{&eee}
\ee

The vectors 
\be
\bigl[{{v_i^{(s)}}_\pm}\bigr]_\rho\equiv\alpha_\pm^i
\bigl(U_{i-1}^{(s_{i-1})}\bigr)^{-1}\dots 
\bigl(U_{1}^{(s_{1})}\bigr)^{-1}{(v^0_\pm)}_\rho,
\label{&vvv}
\ee
where ${(v^0_\pm)}_\rho$ is the $\rho$th component of the vector $v^0_\pm$,
remain collinear for any~$i$ and their ratio $(\alpha_+/\alpha_-)^i
(v_+^0)_\rho/(v_-^0)_\rho$ depends only on the level~$i$. Therefore, it is 
natural to relate the data $(v_+^0)_\rho$ and $(v_-^0)_\rho$ with the
scattering matrix data on the graph~$T$. For this, we call solution
(\ref{&*3}) with only $v_+^0$ nonzero the {\em incoming wave} and 
solution (\ref{&*3}) with only $v_-^0$ nonzero---the {\em outgoing wave}.

Here it becomes clear why we can relate the branch and tail descriptions.
First, we can perform a {\em gauge transformation} in order to eliminate
the dependence on potentials on edges of branches. For this, we merely set
$$
v_i^{(s)}\to U^{(s_1)}_1 U^{(s_2)}_2\dots U^{(s_{i-1})}_{i-1}v_i^{(s)}.
$$
No dependence on potentials on edges remains then in the transformed
variables, and we obtain merely that
$$
\bigl[{v_i}_\pm\bigr]_\rho=\alpha^i_\pm\bigl[{v_0}_\pm\bigr]_\rho
$$
for any path of length~$i$ starting from the summit of the branch.
In what follows, we set $U_{\vec\mu}=I_v$ on all external 
(\ie not entering the minimum reduced subgraph) edges.

We can now identify all vertices and edges that lie at the same distance
from the summit setting 
\be
v^{(i)}_s\to p^{(s-1)/2} v_i \hbox{\ for\ } s>0
\label{&*&}
\ee 
and erasing the
now redundant superscript $(i)$ (see Fig.~3). We thus obtain just the
picture of~\cite{Nov} where a finite (probably zero) number of tails
(half-axes) begin at a vertex of the basis (or, in our notation, reduced)
finite graph. We make transition (\ref{&*&}) in order to preserve
the scalar product (\ref{&brackets}). 
Simultaneously, we must scale and 
shift the corresponding eigenvalue, 
$\lambda\to\bigl[\lambda+\bigl(\sqrt{p}-1\bigr)^2\bigr]/\sqrt{p}$;
only after this operation, the answers for two branches of solution
(\ref{&eee}) in our approach and in the approach of~\cite{Nov} do
coincide.

\phantom{xxx}
{
\setlength{\unitlength}{1pt}

\begin{picture}(50,2)(-80,100)

\multiput(60,20)(100,0){2}{\line(0,1){90}}
\multiput(20,20)(10,0){9}{\circle*{2}}
\multiput(30,50)(30,0){3}{\circle*{3}}
\multiput(60,80)(0,30){2}{\circle*{3}}
\multiput(160,20)(0,30){4}{\circle*{3}}
\put(115,50){\vector(1,0){30}}
\multiput(30,50)(30,0){3}{\line(-1,-3){10}}
\multiput(30,50)(30,0){3}{\line(1,-3){10}}
\multiput(30,50)(60,0){2}{\line(0,-1){30}}
\put(60,80){\line(-1,-1){30}}
\put(60,80){\line(1,-1){30}}
\multiput(66,110)(100,0){2}{\makebox(0,0)[lc]{$v_0$}}
\put(75,85){\makebox(0,0)[cc]{$v_1^{(s_1)}$}}
\put(105,55){\makebox(0,0)[cc]{$v_2^{(s_2)}$}}
\put(115,25){\makebox(0,0)[cc]{$v_3^{(s_3)}$}}
\put(166,80){\makebox(0,0)[lc]{$v_1$}}
\put(166,50){\makebox(0,0)[lc]{$p^{1/2}v_2$}}
\put(166,20){\makebox(0,0)[lc]{$p^1v_3$}}
\end{picture}
}
\vspace{3.5cm}

\centerline{{\bf Fig. 3.}
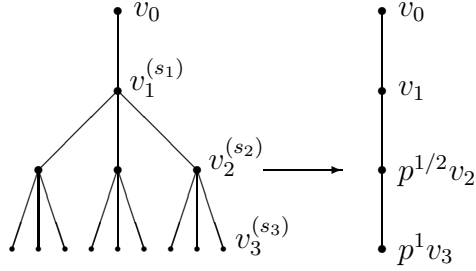
The branch $\to$ tail transition on the external edges of $T$.}

\vspace{5pt}

\begin{defin}{\rm
Let the {\em boundary points} $x_i\in\pa F$ are those points of~$F$ that have
{\em nonzero} number of external tails starting at these points. We also 
consider the natural embeddings of linear spaces $(\pa F)_{0,1}\subset 
\wtd C_{0,1}$ (recall that $(F)_{0,1}\equiv \wtd C_{0,1}$). 
} 
\end{defin}

\subsection{Spherical functions on factorized trees}

We now introduce spherical functions for the graph~$T$. 
First, we consider them on the tree~$D$ itself. Choose the
vertex 
$x\in V(D)$ and claim it the {\it center} of the tree. Then the 
{\em spherical function} $\psi(n,x)$ is an eigenvector of $M_1$,
$M_1\psi(n,x)=t\psi(n,x)$ that depends only on the distance in the tree
from the point $x$, \ie it is constant on each 
sphere $S(n,x)\equiv\{y\in V(D):\, |P_{x,y}|=n\}$.
In the case of trivial potential (trivial representation of 
the group~$\Gamma$), we have
\be 
\psi(n,x)=a_+\alpha_+^n-a_-\alpha_-^n,\label{&4} 
\ee 
where (cf. (\ref{&eee}))
$$
\alpha_+\alpha_-=1/p,
$$ 
and setting $\psi(0,x)=1$, we obtain
\be
a_+=\frac{p\alpha_+-\alpha_-}{(p+1)(\alpha_+-\alpha_-)},\quad
a_-=\frac{p\alpha_--\alpha_+}{(p+1)(\alpha_+-\alpha_-)}.\label{&5}
\ee

In order to define
a similar object for a general graph $T$ whose ``center'' now is the
reduced graph~$F$,
we  consider a {\em superposition} of solutions (\ref{&4})
with the {\em sources} $s_y$ placed at the vertices of $D_F$,
\be
\psi(x)=\sum_{y\in D_F}s_y\psi(d(x,y),y),\label{&5a}
\ee
such that the function $s_y$ is 
periodic w.r.t.\ the action of $\Gamma$: $s_{\gamma y}=s_y$ for all 
$\gamma\in\Gamma$ and $y\in D_F$. 
The behavior of this 
solution on each branch ``growing'' from $\pa F$ is described by (\ref{&4}) 
with the preexponential factors depending only on the point of 
$\pa F$ into which this branch can be projected. 

Let us introduce a kern function $K$:
\be
K(z,y|x)=\sum_{\gamma\in\Gamma}x^{d(z,\gamma(y))}
\times \prod\limits_{\vec \mu_i\in P_{z,\gamma(y)}}^{}U_{\vec\mu_i},\label{&6}
\ee
where $z,y$ are points of $D_F$. This function is 
periodic under the action of $\Gamma$ over both its arguments 
separately. Therefore, it is well-defined on~$T$ itself. 
For the trivial representation $\chi\equiv1$, this 
function is also symmetric in $z$ and $y$.  Function 
(\ref{&5a}) then becomes  
\be 
\psi(x)=\sum_{y\in 
F}s_y\bigl[a_+K(t(x),y|\alpha_+)\alpha_+^{d(x)}- 
a_-K(t(x),y|\alpha_-)\alpha_-^{d(x)}\bigr],\label{&7}
\ee
where $t(x)\in F$ is the image of the 
point~$x\in D$. (If $x\in F$, then $t(x)=x$.)  
We call the part of (\ref{&7}) proportional to $\alpha_+^{d(x)}$
the {\em retarded} wave function and the part proportional to
$\alpha_-^{d(x)}$ the {\em advanced} wave function.  
Therefore, $\psi(x)$ has a general form 
\be 
\psi(x)=A_{adv}(t(x))\alpha_+^{d(x)}-
A_{ret}(t(x))\alpha_-^{d(x)}\equiv \psi_+(x)-\psi_-(x),
\label{&77}
\ee

\begin{prop}\label{&XX}
Given a set of vectors $v_+^0(x_i)$, \ $x_i\in\pa F$, the set of $v_-^0(x_i)$
is uniquely determined for all $t\in {\bf C}$ except a finite set of
points {\rm(}the points of the {\em discrete spectrum}{\rm)}. 
\end{prop}

The important case of functions (\ref{&77}) is where $\psi_+(x)$ is nonzero
on only one branch growing from a single point of $\pa F$. Then, we can
define the scattering matrix as follows.

\begin{defin}{\rm
The scattering matrix
($S$-matrix) $S(t)$ is the square matrix of the size 
$\hbox{rank\,}(\pa F)_0$ with the entries $s_{i_\alpha,j_\beta}$, \
$\bigl|\{i_\alpha\}\bigr|=\bigl|\{j_\beta\}\bigr|=\hbox{rank\,}(\pa F)_0$,
\ $s_{i_\alpha,j_\beta}$ is the value of $\bigl(v_-^0\bigr)_{\beta}(x_j)$
for $\bigl(v_+^0\bigr)_{\gamma}(x_k)\equiv \delta_{\alpha\gamma}\delta_{ki}$
and $\bigl(v_\pm^0\bigr)$ pertain to the solution of spectral problem 
(\ref{&*}) on the whole graph, \ie entries of the $S$-matrix correspond to
solutions of (\ref{&*}) of the form
\be
\psi_{i_\alpha}^+ -\sum_{\{j_\beta\}}^{}s_{i_\alpha,j_\beta}\psi_{j_\beta}^-.
\label{&777}
\ee
}
\end{defin}

We postpone the detailed study of the spectral properties of 
the $S$-matrix to the next section and formulate here the main theorem
connecting $S$-matrix and determinants of the operators 
on the reduced graph.

\begin{theorem}\label{&th2}
In the case of unitary potentials on edges,
$$
\det S(t)=\left(\frac{\alpha_+}{\alpha_-}\right)^{\hbox{\scriptsize
rank\,}(F)_0}
\frac{\det\bigl(\Delta_\rho(\alpha_-)\bigr)}
{\det\bigl(\Delta_\rho(\alpha_+)\bigr)},
$$
where $\bigl(\Delta_\rho(\alpha_+)\bigr)$ is operator {\rm(\ref{&D})}
in which we explicitly indicate the dependence on the representation~$\rho$.
\end{theorem}

{\em Proof.} Each solution of spectral problem (\ref{&*}) having form 
(\ref{&*3}) outside the reduced graph can be split into the advanced and
retarded partial wave functions in the total graph (see (\ref{&77})) where
\be
\psi_\pm(x)=\sum_{x_\alpha\in F}^{}\sum_{P_{x_\alpha,x}}^{}
a_\pm\,\alpha_\pm^{|P_{x_\alpha,x}|}\prod_{\vec e_i\in P_{x_\alpha,x}}^{}
U_{\vec e_i}^{-1}s(x_\alpha).
\label{&*5}
\ee
Here $s(x_\alpha):F_0\to V$ is the {\em source function} and
$a_\pm$ are from (\ref{&5}).

Note that the ``splitting'' into the advanced and retarded parts or, 
equivalently, the source function $s(x_\alpha)$ is inambiguously
defined everywhere outside $F$ (the splitting is also
inambiguously defined at boundary points
$x\in \pa F$), but is ambiguous at the 
inner points of~$F$. 

In the space of $s(x)$ (in fact, this space is $\wtd C_0$),
there exists a linear subspace of source functions that
generate one and the 
same eigenfunction $\psi(x)$. For example, if
$x_0\in F$, \ $x_0\not\in \pa F$, we can propose two source functions
$s_1(x)=\{0,x\ne x_0;\ 1,x=x_0\}$ and 
$s_2(x)=\{0,d(x,x_0)\ne1;\ 2p/t,d(x,x_0)=1\}$. Then, these two source 
functions generate the {\em same} eigenfunction $\psi(x)$. Moreover, for 
almost all~$\lambda$ and $s(x)\in \wtd C_0$, 
there exists a {\em unique} $\bar s(x)\in \wtd C_0$ such that both
$s(x)$ and $\bar s(x)$
generate the same $\psi(x)$ and $\bar s(x)\equiv 0$ in all 
inner points of~$F$.

Therefore, the problem is to define a ``convenient'' 
splitting (\ref{&77}) (an analogue of a finite-dimensional ``gauge
fixing'' in field theory). 

Let us consider {\em eigenvalues} $c_n$ of $S(t)$. Those are numbers for which
there exist solutions of (\ref{&*}) with the sets $(v_+^0)_\alpha(x_i)$
and $(v_-^0)_\alpha(x_i)$ {\em proportional} to each other:
\be
(v_+^0)_\alpha(x_i)=c_n(v_-^0)_\alpha(x_i)\quad\hbox{for\ all\ $\alpha$\
and\ $x_i\in \pa F$}.
\label{&*4}
\ee
We now expand the condition (\ref{&*4}) to {\em all} 
(external as well as internal) points of~$F$.
\begin{defin}\label{&defSp}{\rm
Let the {\em spherical function\/} on the graph~$T$ be the solution of
spectral problem (\ref{&*}) of form (\ref{&*5}) such that
\be
\bigl[\psi_+(x)\bigr]_\alpha/\bigl[\psi_-(x)\bigr]_\alpha=
c_n\left(\frac{\alpha_+}{\alpha_-}\right)^{\hbox{\scriptsize dist\,}(x,F)},
\label{&*6}
\ee
where $c_n$ is the constant independent on the point~$x$ and the 
representation index~$\alpha$ and $\hbox{dist\,}(x,F)$ is the 
well-defined distance between the point~$x$ and the reduced graph~$F$ (for 
$x\in F$, \ $\hbox{dist\,}(x,F)=0$).
}
\end{defin}

\begin{lemma}\label{&lem2}
$$
\det S(t)=\prod_{n}^{}c_n.
$$
\end{lemma}

Actually,
condition (\ref{&*6}) fixes the choice of the admitted source
functions $s(x_\alpha)$. At the same time,
this is a system of linear homogeneous
equations, which has nonzero solution only for a finite set of $c_n$.
Below, we present this system in terms of the Hecke operators 
(\ref{&Mn1}) on the graph~$F$.

Considering Eq.~(\ref{&*6}) at the points of the reduced graph, 
we obtain
$$
\sum_{x_\alpha\in F}^{}a_+\,\alpha_+^{|P_{x_\alpha,x}|}
\left(\prod_{\vec e_i\in P_{x_\alpha,x}}^{}U_{\vec e_i}^{-1}\right)
s(x_\alpha) =
c_n \sum_{x_\alpha\in F}^{}a_-\,\alpha_-^{|P_{x_\alpha,x}|}
\left(\prod_{\vec e_i\in P_{x_\alpha,x}}^{}U_{\vec e_i}^{-1}\right)
s(x_\alpha),
$$
or
\be
\sum_{y\in F}s_ya_+K(z,y|\alpha_+)=c_n\sum_{y\in
F}s_ya_-K(z,y|\alpha_-).
\label{&8}
\ee
Then, obviously, 
\be
\prod_{n}^{}c_n=\frac{\det K(x,y|\alpha_+)}{\det K(x,y|\alpha_-)}
\left(\frac{a_+}{a_-}\right)^{\hbox{\scriptsize rank\,}(F)_0},
\label{&*10}
\ee
where $K(x,y|\alpha)$ is the kernel for the linear operator acting on
$\wtd C_0$ (note that because~$F$ contains all loops of the 
graph~$T$ and is connected, then, if
$x,y\in F$, then any path $P_{x,y}\subseteq F$):
$$
(Ks)(x)=\sum_{y\in F}^{}\sum_{P_{x,y}}^{}\alpha^{|P_{x,y}|}
\left(\prod_{\vec e_i\in P_{x,y}}^{}U_{\vec e_i}^{-1}\right)s(y).
$$
Considering the universal covering $D_F$, it is easy to verify that
$$
\tilde\Delta(\alpha)K(x,y|\alpha)=(1-\alpha^2)I_0,
$$
and, therefore,
$$
\det K(x,y|\alpha)=(1-\alpha^2)^{r_0}
\det{}^{-1}\tilde\Delta(\alpha).
$$
Now, using the identity
$$
\frac{a_+(1-\alpha^2_+)}{a_-(1-\alpha^2_-)}=\frac{\alpha_+}{\alpha_-}
$$
and formula (\ref{&*10}), we obtain the assertion of Theorem \ref{&th2}.

Theorems \ref{&th1} and \ref{&th2} imply the following correspondence
between the scattering matrix determinant and the $L$-function of the
{\em reduced graph\/}:
\be
\det S(t)=\left(\frac{\alpha_+}{\alpha_-}\right)^{r_0}
\left(\frac{1-\alpha_-^2}{1-\alpha_+^2}\right)^{r_0-r_1}
\frac{L(\rho,\alpha_+)}{L(\rho,\alpha_-)},\quad \alpha_+\alpha_-=1/p,
\label{&*11}
\ee
where $L(\rho,\alpha)$ depends only on characters of loops of the
reduced graph~$F$. Another form of (\ref{&*11}) is
\be
\det S(t)=\left(\frac{\alpha_+}{\alpha_-}\right)^{r_0}
\frac{\det\tilde\Delta(\alpha_-)}{\det\tilde\Delta(\alpha_+)}.
\label{&*12}
\ee

Note that only the first ``volume factor'' in (\ref{&*11}) and (\ref{&*12})
depends on the volume of the 
graph~$F$, all nontrivial factors depend only on loop characteristics of 
$\Tr$.

Since $\alpha_-\alpha_+=1/p$, the combination 
$(\alpha_\pm)^{r_0}\det\tilde\Delta(\alpha_\mp)$ acquires the form
\be
(\alpha_\pm)^{r_0}\det\tilde\Delta(\alpha_\mp)= 
\det\bigl(\alpha_\pm\tilde\Delta((\alpha_\mp)\bigr) 
=\det\left(\alpha_\pm+\frac{\alpha_\mp}{p}\t Q-\frac 1p\t M_1\right).
\label{&4.23}
\ee

\begin{remark}\label{&rem10}{\rm
Returning to (\ref{&8}), we see that the determinant of the 
operator $K(z,y|\alpha_\pm)$ is nondegenerate (everywhere except 
discrete points of spectrum of $\tilde\Delta(u)$). It is in contrast to the 
already mentioned ambiguity in choice of the source function $s(x)$. 
Let us now act by the operator $\Delta(\alpha_+)$ on both sides of 
(\ref{&8}). For any interior point of $F$,
$$
\bigl(\alpha_-\Delta(\alpha_+)s\bigr)(x)=
\bigl(\alpha_+\Delta(\alpha_-)s\bigr)(x);
$$
therefore, we obtain $s(y)=c_is(y)$, \ie $s(y)\equiv0$ in all internal 
points for all $c_i\ne 1$. Hence, using condition (\ref{&8})
we gauge out the whole ambiguity 
due to the choice of $s(y)$ at almost all values of the
spectral parameter~$t$. (The cases where such a condition fails just
correspond to the exceptional points of the discrete spectrum, see below.)
}
\end{remark}

\subsection{Structure of spectrum}

We now study the spectral properties of problem (\ref{&*}). These properties
turn out to be quite similar to the spectral properties of Schr\"odinger
type potentials investigated in~\cite{Nov}.

We now consider
eigenfunction problem (\ref{&*}) as the spectral problem. 

\begin{defin}\label{&d_spec}
{\rm
The {\it continuous spectrum}, or the {\it scattering zone of the spectrum},
is the domain $t^2<1/p$ where $\alpha_+$ and $\alpha_-$ are complex 
conjugate to each other and their absolute value is exactly $1/\sqrt{p}$;
then, 
\be
\lambda\in\bigl(-p-1-2\sqrt{p},-p-1+2\sqrt{p}\bigr).
\label{&cont}
\ee

The points of {\it normal discrete spectrum} are eigenvalues of problem
(\ref{&*}) that lie outside the continuous spectrum domain (\ref{&cont})
and correspond to real solutions that decay exponentially (as $\alpha_+^n$)
with $|\alpha_+|<1/\sqrt{p}$ and such that these eigenfunctions are
nonzero at least in one branch.

The points of the {\it exceptional discrete spectrum} may appear for any real
$\lambda$. Those are points such that the corresponding eigenfunctions
{\it vanish identically} outside the reduced graph.
}
\end{defin}

The first two types of points are customary for a majority of spectral 
problems. The third type is rather specific for graph problems 
or, more generally, for discrete spaces (its
existence in the problem under consideration was observed in~\cite{Roman}).

Note that points of the exceptional discrete spectrum exist not for all 
graphs; customarily, it is some graph symmetry (e.g., the ${\Bbb Z}_2$
symmetry, see~\cite{Nov}) that is responsible for the appearance of
such points. Obviously, for such an exceptional solution to exist,
it must vanish not only on all branches, but also at all summits of these
branches, \ie at {\it all} boundary points $x\in\pa F$ (and,
correspondingly, at all boundary points of the minimum reduced graph).
This condition is not easy to formalize; 
however, in the unitary potential case, we can formulate
some regular
criterion for the existence of exceptional discrete points in spectrum
(see Proposition \ref{&propEx}).

If $k$ is the total number of branches growing from the points of $\pa F$,
then, for any $\lambda$ from the scattering zone, considering the
spherical function case (the functions
that are constant on spherical slices of branches)
we always have a $k$-dimensional subspace of eigenfunctions corresponding
to the given eigenvalue~$\lambda$. 

Fixing the spectral parameter~$\lambda$,
we can introduce the $2k$-dimensional space $H^{2k}$
of {\it asymptotic states}---the asymptotic spherical function-like
local solutions of (\ref{&*}), which always has two independent
solutions proportional to $\alpha_+^n$ and $\alpha_-^n$ in each branch.
These functions, $\psi^\pm_i$, \ $i=1,\dots,k$, constitute a basis 
in $H^{2k}$..
Then, the true scattering problem solutions (\ref{&777}), \ie those
that can be
continued to the whole graph, span a $k$-dimensional subspace 
$\Lambda^k$ of $H^{2k}$.

\begin{prob}\label{&prob1}
{\rm
Is it possible to develop a Lagrangian plane description for the
unitary potential case in analogy with the
Schr\"odinger potential case~\cite{Nov}?
}
\end{prob}

\begin{prop}\label{&propS} The $S$-matrix 
{\rm(\ref{&777})} is unitary in the scattering domain {\rm(\ref{&cont})}.
\end{prop}
{\it Proof.} The proof resembles (however, with some
important variations) the proof~\cite{Nov} of the analogous
statement for the Schr\"odinger potential scattering case.

As the first step, we define the symmetric function $W(\phi,\psi)$
(we call this function s-Wronskian by
analogy with the {\it antisymmetric} Wronskian in~\cite{Nov}), which sets
into the correspondence to two functions from $C_0$ the function from
$C_1$ by the formula
\be
W(\phi,\psi)=\sum_{\vec \mu_{x,y}\in \vec L}^{}\bigl(\phi^\ast(x) 
U_{\vec \mu_{x,y}}\psi(y)+
\psi^\ast(x) U_{\vec \mu_{x,y}}\phi(y)\bigr).
\label{&Wron}
\ee
If $\phi$ and $\psi$ are two eigenfunctions from $\Lambda^k$ 
(by definition corresponding to the same 
value of the spectral parameter~$\lambda$), then
the s-Wronskian is a chain (one-cycle with possible open ends), \ie in
the infinite graph, we have
$$
\pa W(\phi, \psi)=0.
$$
This follows from the simple calculation,
\bea
\pa W(\phi, \psi)(x)&=&
\phi^\ast(x)\sum_{y}^{}U_{\vec\mu_{x,y}}\psi(y)
+\psi^\ast(x)\sum_{y}^{}U_{\vec\mu_{x,y}}\phi(y)-\nonumber\\
&{}&-\phi^\ast(y)\sum_{y}^{}U_{\vec\mu_{y,x}}\psi(x)
-\psi^\ast(x)\sum_{y}^{}U_{\vec\mu_{y,x}}\phi(x)=\nonumber\\
&=&\phi^\ast(x)(M_1\psi)(x)
-\bigl(\phi^\ast(x)(M_1\psi)(x)\bigr)^\ast+\nonumber\\
&{}&+\psi^\ast(x)(M_1\phi)(x)
-\bigl(\psi^\ast(x)(M_1\phi)(x)\bigr)^\ast.
\label{&s-W}
\eea
Using now the relation $M_1\phi=(Q+1-\lambda)\phi$ and remembering that
$Q$ and $\lambda$ are Hermitian, we obtain zero in the r.h.s.\ 
of (\ref{&s-W}).

The s-Wronskian of two solutions is therefore a chain. This means in turn 
that if we now cut the branches of the (infinite) graph at some distance
$n$ from the reduced graph~$F$ (this distance must not be the same for all
branches; it is only preferable to be the same on each branch), 
thus obtaining some finite graph $T_{\CUT}$, 
then, nevertheless, the {\it total} sum vanishes:
\be
\sum_{x\in \pa T_{\CUT}}^{}\pa W=0.
\label{&boundary} 
\ee
Here the sum ranges all boundary points 
$x\in \pa T_{\CUT}$, and $\pa W$ is restricted to the graph
$T_{\CUT}$.

It is convenient to introduce the bilinear pairing on $H^{2k}$,
$$
\<\<\psi,\phi\>\>\equiv \sum_{x\in \pa T_{\CUT}}^{}\pa W(\psi,\phi)
$$ 
on the {\it boundary} of $T_{\CUT}$ for two arbitrary
asymptotic vectors from $H^{2k}$.

The reasoning in the beginning of Sec.~4 show that we can 
consistently eliminate
the gauge potential dependence on all external branches. Then,
the $\psi^\pm_i$ functions are mere exponents of $\alpha_\pm$: \
$\psi^\pm_i=\alpha_\pm^n$ on $i$th branch and zero otherwise.
Then, using the identities $\alpha_+\alpha_-=1/p$ and 
$\alpha_-=\alpha_+^\ast$, which hold in the scattering zone, 
it is easy to find that
$$
\<\<\psi^\pm_i,\psi^\pm_j\>\>=\pm\delta_{i,j}\frac{\alpha_--\alpha_+}{2p}
\equiv\pm\delta_{i,j}\<\<\psi^+,\psi^+\>\>
$$
and
$$
\<\<\psi^\pm_i,\psi^\mp_j\>\>\equiv 0.
$$
Now we remember that for two arbitrary solutions $\psi$ and $\phi$
of spectral problem (\ref{&*}) determined
on the whole graph, the expression for $\pa W(\psi,\phi)$
exactly coincides with $\<\<\psi,\phi\>\>$, where only asymptotic
states must be taken into account. Consider now the eigenfunctions
$\Psi_i\equiv \psi_{i}^+ -\sum_{j}^{}s_{i,j}\psi_{j}^-$
\ (cf. formula (\ref{&777})).\footnote{We
use condensed multiindex notation.} 
Then, for two such functions, we have
\be
0=\<\<\Psi_i,\Psi_j\>\>\equiv\<\<\psi^+,\psi^+\>\>\biggl(
\delta_{i,j}-\sum_{k}^{}s^\ast_{i,k}s_{j,k}\biggr).
\label{&7777}
\ee
Because $\<\<\psi^+,\psi^+\>\>\ne0$, we obtain that the $S$-matrix 
is unitary. The proposition is therefore proved.\footnote{Considering
formula (\ref{&*11}) we obtain that the 
$S$-matrix determinant is a unitary function as soon as $\alpha_\pm$ lie
in the scattering zone; this is of course
in accordance with Proposition \ref{&propS}.}

It is interesting that we can deduce the existence of the exceptional
discrete spectrum from the $L$-function itself.

\begin{prop}\label{&propEx}
Exceptional discrete spectrum appears iff $\exists\alpha_+,\alpha_-:$ \ 
$\alpha_+\alpha_-=1/p$ and the $L$-function has poles at 
both $\alpha_+$ and $\alpha_-$.
\end{prop}

{\it Sketch of the proof.} We seek the exponential solution of 
spectral problem (\ref{&*}) such that one of the waves, retarded
or advanced, just vanish identically in all vertices of the graph~$T$
(including now also {\it internal} points of~$F$). This immediately implies
that there exists a vector (source function) $f_0\in\wtd C_0$ such that
$$
\Delta(\alpha_+)f_0=0.
$$
Then the nonlocal operator $\widehat K$,
$$
\widehat K f(x)=\sum_{y\in F}^{}K(x,y|\alpha)f(y),\quad f\in \wtd C_0,
$$
which is inverse to $\tilde\Delta(\alpha)$, becomes ill defined.

Let $f_0$ be the null vector of $\tilde\Delta(\alpha)$. This immediately
produces the characteristic equation on~$\alpha$:
$$
\det \tilde\Delta(\alpha_+)=0,
$$
and if simultaneously 
$\tilde\Delta(\alpha_-)\ne 0$, then, following Theorem~\ref{&th2}, 
the $S$-matrix develops a singularity (zero or pole), which indicate that
for such value of~$\lambda$ there exists a solution of spectral problem
(\ref{&*}) in which one (and only one) of the retarded and advanced
waves is nonzero (cf.~\cite{Roman}).

Thus, if no cancellations between numerator and denominator occur in 
formula (\ref{&*11}), \ie the assertion of the proposition is not true, then
the discrete spectrum is exhausted by regular eigenfunctions and no
exceptional spectrum exists. 

On the contrary, when there are such poles of the $L$-function that
the assertion of the proposition is satisfied, then the regular
discrete spectrum does not provide {\it all} possible solutions of the
characteristic equation; the solutions that yield zero both in the denominator
and in the numerator of (\ref{&*11}) must therefore correspond to the
exceptional discrete eigenvalues. 

\begin{prob}
{\rm
To make this sketch of the proof the formal proof.
}
\end{prob}

\newsection{Examples. Algorithm for calculating $L(1,u)$.}

For simplicity, we assume in what follows that $F=D(\Gamma)/\Gamma$.
We omit most of the proofs that are simple exercises in linear algebra.

\begin{example}\label{&*ex1}{\rm
{\em Graph $T$ with no external branches.} 
Here we consider a limiting case of our construction when the graph $T$
exactly coincides with $\Tr$. In this 
no-scattering case, $q(z)\equiv p$ and we obtain
the inversion relation for $\Du$. If $\alpha_-\alpha_+=1/p$, 
we have 
\bea
\det{}^{-1}\bigl[(1+p\alpha_+^2)I-\alpha_+M_1\bigr]&=&
\det{}^{-1}\left[1+\frac{1}{p\alpha_-^2}-\frac{1}{p\alpha_-}M_1\right]
\nonumber\\
&=&\bigl(\alpha_-^2p\bigr)^{r_0}
\det{}^{-1}\bigl[(1+p\alpha_-^2)I-\alpha_-M_1\bigr]\nonumber\\
&=&\frac{\alpha_-^{r_0}}{\alpha_+^{r_0}}
\det{}^{-1}\bigl[(1+p\alpha_-^2)I-\alpha_-M_1\bigr],\label{&inverse}
\eea
and using (\ref{&*12}) and (\ref{&inverse}), we obtain that the
total scattering matrix $C\equiv 1$.  Therefore,
$C$ is trivial and does not depend at all on the shape of 
the reduced (or, in this case, total) graph $T=\Tr$. This is in accordance
with the fact that {\it all} spectral points are points of the exceptional
discrete spectrum in this case.
}
\end{example}

\begin{example}\label{&*ex2}
{\rm {\em A one-loop case} ($g$=1). In the case where the graph~$T$
contains a single loop,
the $L$-function in the case of the trivial character
$\chi(u)=1$ is
\be
L(1,u)_{g=1}=\frac 1{(1-u^n)^2},\label{&L1}
\ee
because the group $\Gamma_1$
contains only two primitive elements 
$\gamma$ and $\gamma^{-1}$ with the same length
$l(\gamma)=l(\gamma^{-1})=n$.
}
\end{example}

\subsection{Two-loop case. The general structure of $L(1,u)$.}

We present an algorithm for calculating $L(u)\equiv L(1,u)$ for 
arbitrary graph. Let $l_i$ be the lengths of edges of some 
graph $\Tr$ and $a_i\equiv u^{l_i}$. 
Then, from Theorem~1, we have $L^{-1}(u) 
=\det(1-uT_1)=P_{2r_1}$---the polynomial of degree $2\sum l_i$
in the variable~$u$. 
This polynomial can be obviously treated as a polynomial in $a_i$ 
(all closed geodesics are built from intervals connecting three- and more-
valent vertices of
$\Tr$) and we may treat $a_i$ as formal independent variables.
Assuming that one of $l_i$ is greater than the sum of all others,
we conclude that the maximum degree of the polynomial 
$P_{2r_1}(u)$ in each $a_i$ is less or equal $2$
Thus, $L^{-1}(u)=P_2(\{a_i\})$. This polynomial has integer coefficients 
and from the relation
$$
L^{-1}(u)=\det (1-uT_1)= {(1-u^2)}^{g-1}\det \Du
$$
it follows that $L^{-1}(u)$ has zero of order at least $g-1$ at the 
point $u=-1$ and zero of exact order~$g$ at $u=1$. This 
follows from the relation $\bigl.\det\Du\bigr|_{u=1}=0$,
which holds for arbitrary graph, and from the relation
$$
\frac{\pa}{\pa u}\bigl.\det\Du\bigr|_{u=1}=\hbox{\#\ maximum\ trees},
$$
(see~\cite{Ch3} where it was noted that the derivative of $\Du$ at $u=1$
is expressed by the Kirchhoff formula 
through the total number of maximum connected 
trees in the graph~$F$ and is therefore always nonzero).

We turn now to two-loop case. There we have three possibilities 
(see~Fig.~4, a, b, and~c).

\eop

\phantom{XXX}

\begin{picture}(190,2)(-20,85)

\put(20,50){\oval(40,40)}
\put(20,30){\line(0,1){40}}
\multiput(20,30)(0,40){2}{\circle*{3}}
\put(10,45){\makebox{$l_1$}}
\put(25,45){\makebox{$l_2$}}
\put(45,45){\makebox{$l_3$}}
\put(15,15){\makebox{a}}
\multiput(80,50)(60,0){2}{\oval(30,30)}
\put(95,50){\line(1,0){30}}
\multiput(95,50)(30,0){2}{\circle*{3}}
\put(75,70){\makebox{$l_1$}}
\put(105,55){\makebox{$l_2$}}
\put(135,70){\makebox{$l_3$}}
\put(105,15){\makebox{b}}
\multiput(185,50)(30,0){2}{\oval(30,30)}
\put(200,50){\circle*{3}}
\put(180,70){\makebox{$l_1$}}
\put(210,70){\makebox{$l_2$}}
\put(195,15){\makebox{c}}
\end{picture}

\vspace{4.cm}

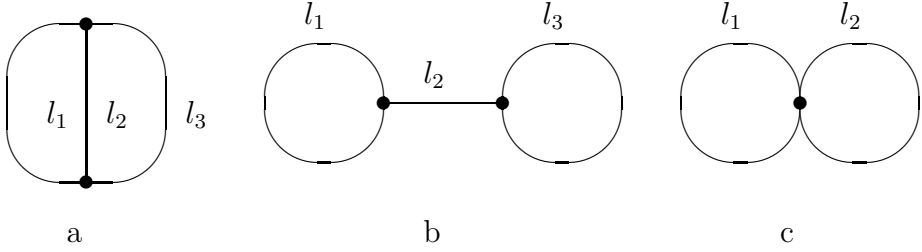
\centerline{{\bf Fig. 4.} Three possible graphs for $g=2$.}

\vspace{6pt}

The analysis in~\cite{Ch2} shows how one can easily calculate the
corresponding $L$-functions from the above considerations. The answer is
\bea
\hbox{case\ a}\quad L^{-1}_{(a)}&=&\left(1-\sum_{i<j}a_ia_j 
-2a_1a_2a_3\right)\left(1-\sum_{i<j}a_ia_j +2a_1a_2a_3\right),\nonumber\\
\hbox{case\ b}\quad L^{-1}_{(b)}&=&(1-a_1)(1-a_2) 
\bigl[(1-a_1)(1-a_2) -4a_3^2a_1a_2\bigr],\nonumber\\
\hbox{case\ c}\quad L^{-1}_{(c)}&=&(1-a_1)(1-a_2) 
\bigl[(1-a_1)(1-a_2) -4a_1a_2\bigr].
\eea
(Note that the case~c can be obtained either from~a or from~b 
case by setting $a_3=1$.) 

\begin{example}\label{&*ex3} {\rm
{\it Comparing the Hermitian and unitary potentials.}
We now consider $L$-func\-tions and the corresponding determinants
for two graphs depicted in Fig.~5 endowed with Abelian potentials.

\begin{picture}(190,2)(-40,85)
\put(20,50){\oval(40,40)}
\put(20,30){\line(0,1){40}}
\multiput(20,30)(0,40){2}{\circle*{3}}
\put(5,45){\makebox{$a_1$}}
\put(25,45){\makebox{$a_2$}}
\put(45,45){\makebox{$a_3$}}
\multiput(0,50)(20,0){3}{\vector(0,1){0}}
\put(15,15){\makebox{a}}
\put(100,50){\oval(40,40)}
\put(100,30){\line(0,1){40}}
\multiput(100,30)(0,40){2}{\circle*{3}}
\put(80,50){\circle*{3}}
\put(86,36){\makebox{$c$}}
\put(86,59){\makebox{$c$}}
\put(105,45){\makebox{$a$}}
\put(125,45){\makebox{$a^{-1}$}}
\multiput(100,50)(20,0){2}{\vector(0,1){0}}
\put(84,34){\vector(-1,1){0}}
\put(84,66){\vector(-1,-1){0}}
\put(95,15){\makebox{b}}
\end{picture}

\vspace{4.cm}

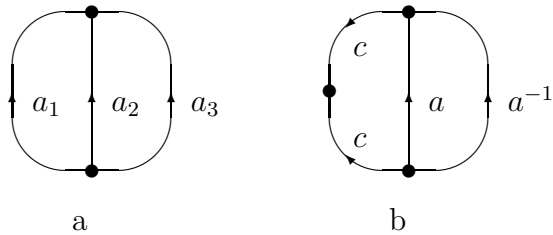
\centerline{{\bf Fig. 5.} Examples of reduced
graphs with Hermitian and unitary
potentials.}

\vspace{6pt}

We present answers for Hermitian and unitary potentials (where 
$a_i\to a_i^{-1}$ for the inversely oriented edges) in case~a.
For the Hermitian potential, we obtain
$$
\det \Delta(u)=(1+2u^2)^2-s^2u^2, \quad s\equiv a+b+c,
$$
$$
\det (1-T(u))=1-2u^2(ab+bc+ca)+u^4(abb+bc+ca)^2-4u^6a^2b^2c^2,
$$
while for the unitary potential, these quantities are, of course,
related:
$$
\det \Delta(u)=(1+2u^2)^2-s\tilde su^2, \quad s\equiv a+b+c,\ 
\tilde s\equiv a^{-1}+b^{-1}+c^{-1},
$$
$$
\det(1-T(u))=(1-u^2)\det \Delta(u).
$$
Case b in Fig.~5 is of interest because it is just the case where
the exceptional discrete spectrum is nonempty for $p=2$ and for any~$a$
(it suffices to set equal in the absolute value
positive and negative charges at the upper and lower vertices
in Fig.~5b; then, the eigenfunction 
vanishes identically at the middle point).
For the $L$-function, we obtain
$$
L(u)=(1+ux+2u^2)(1-ux+u^2-u^3x+2u^4),\quad x\equiv a+a^{-1},\ p=2,
$$
\ie the condition of Proposition \ref{&propEx} is satisfied
for any~$a$, because
the product of two roots of the first quadratic polynomial is $1/2\equiv 1/p$
in this case.
}
\end{example}

\newsection{Teichm\"uller spaces via graphs}

In this section, we consider 
Teichm\"uller spaces ${\cal T}^h$---the spaces of 
complex structures on (possibly open) Riemann surfaces~$S$ with holes
(punctures)
modulo diffeomorphisms homotopy equivalent to identity.
In the vicinity of a boundary component, the complex structure 
can be isomorphic as a complex manifold 
either to an annulus (hole) or to a punctured disc (puncture). 

The graph description following~\cite{Penner} and~\cite{Fock}
is suitable for
considering the finite covering ${\cal T}^H(S)$
of the Teichm\"uller space ${\cal T}^h(S)$. 
A point of ${\cal T}^H(S)$ is 
determined by a point of ${\cal T}^h(S)$ and by the orientation 
of all holes of~$S$ that are not punctures. 
(This covering is obviously ramified over the 
subspace of surfaces with punctures.)

It is well known that an oriented 2D surface with negative Euler 
characteristic can be continuously
conformally transformed to the constant curvature surface.
The Poincar\'e uniformization theorem claims that any complex
surface $S$ of a constant negative curvature (equal $-1$ in what
follows) is a quotient of the upper half-plane ${\HH}_+$
endowed with the hyperbolic metric $ds^2=dzd\overline z/(\Im z)^2$
over the action of a discrete Fuchsian subgroup~$\Delta(S)$
of the automorphism group $PSL(2,\RR)$,
$$
S={\HH}_+/\Delta(S).
$$
In the hyperbolic metric, geodesics are either
half circles with endpoints
at the real line $\RR$ or vertical half-lines; all points of the
boundary $\RR$ are at infinite distance from each other and from any
interior point.

Any hyperbolic homotopy class of closed curves $\gamma$
contains a unique {\it closed
geodesic} of the length $l(\gamma)=\left|\log
{\lambda_1}/{\lambda_2}\right|$,
where $\lambda_1$ and $\lambda_2$ are (different)
eigenvalues of the element of $PSL(2,\RR)$ that corresponds to~${\gamma}$.

Since Strebel \cite{Streb1}, the fat, or ribbon, graphs have been used to 
coordinatize the Teichm\" uller and moduli space. We use a 
rather explicit and simple version of this description~\cite{Fock}.

\begin{claim}\label{&cl1}
%{\rm (Strebel~\cite{Streb1}).}
For a given three-valent fat graph $T$
of genus~$g$ and number of punctures~$n$,
there exists a one-to-one
correspondence between the
set of points of ${\cal T}^H(S)$ and the set
${\bf R}^{\hbox{\small \#\,edges}}$
of edges of this
graph supplied with real numbers
{\rm(}lengths{\rm)}.
\end{claim}

We propose the explicit way how to construct the
Fuchsian group $\Delta(S)\subset PSL(2,\RR)$, which
corresponds to a given set of numbers 
on edges of a graph $T \in T(S)$
such that $S=\HH_+/\Delta(S)$.\footnote{
Note that $\pi_1(S)$ is isomorphic to $\pi_1(\Gamma)$.}
For this, we 
must associate an element $P_\gamma\in PSL(2,\RR)$ to any element of the 
fundamental group $\gamma \in \pi_1(S)$.

To each edge~$\alpha$ we
associate the matrix $X_{z_\alpha}\in PSL(2,\RR)$ of the M\"obius 
transformation
\be
\label{&XZ}
X_{z_\alpha}=\left(
\begin{array}{cc} 0 & -\e^{Z_\alpha/2}\\
                \e^{-Z_\alpha/2} & 0\end{array}\right).
\ee
In order to parameterize a {\it path} over edges of the graph, we introduce
the matrices of the ``right'' and ``left'' turns
\be
\label{&R}
R=\left(\begin{array}{cc} 1 & 1\\ -1 & 0\end{array}\right), \qquad
L\equiv R^2=\left(\begin{array}{cc} 0 & 1\\ -1 &
-1\end{array}\right).
\ee
The spaces $C_0$ and $C_1$ are spaces of functions that take values in the
fundamental two-dimensional representation of the group $PSL(2,{\RR})$.

The ``operators of the right and left turns,''
$R_z$ and $L_z$, are
\bea
\label{&Rz}
R_z\equiv RX_z&=&\left(\begin{array}{cc}
                \e^{-Z/2}&-\e^{Z/2}\\
                     0   &\e^{Z/2}
                     \end{array}\right),\\
\label{&Lz}
L_z\equiv LX_z&=&\left(\begin{array}{cc}
                \e^{-Z/2}&   0\\
                 -\e^{-Z/2}&\e^{Z/2}
                     \end{array}\right).
\eea

The operator $T(u)$ acts on the
space $C_1$ as follows.
Let ${\vec e}_R$ (respectively, ${\vec e}_L$) be the oriented edge
to which the oriented edge ${\vec e}_z$ is naturally mapped by the
right (respectively, left) turn when going along consecutive
edges of a path. For $u\in {\Bbb C}$  and
${\vec e}\in \vec L$, we obtain
$$
T(u)v_{\vec e}\cdot{\vec e}_z=v_{{\vec e}_R}\cdot{\vec e}_R +
v_{{\vec e}_L}\cdot{\vec e}_L,
$$
where
$$
v_{{\vec e}_R}=uR_zv_{\vec e}\ \hbox{and}\
v_{{\vec e}_L}=uL_zv_{\vec e}.
$$

A {\it geodesic} is a {\it closed primitive path} in the graph~$T$.
To each such path we set into the
correspondence the product of matrices
$P_{z_1\cdots z_n}=
L_{z_n}L_{z_{n-1}}R_{z_{n-2}}\cdots R_{z_2}L_{z_1}$, where the
matrices $L_{z_i}$ or $R_{z_i}$ are inserted depending on
which turn---left or right---the path is going on the corresponding step.

The matrices~$L$ and~$R$ are torsion potentials from Definition~\ref{&def3}.
However, we can modify the original graph~$T$
at each vertex disconnecting
edges at the vertex and connecting them by three ``short'' edges
forming a triangle subsequently erasing one (any) of the new edges
in order to preserve the number of loops of the graph (see Fig.~6). 

{\setlength{\unitlength}{1.3mm}%
\begin{picture}(50,27)(-0,48)
\thicklines
\thinlines
\multiput(27,60)(40,0){2}{\vector( 1, 0){ 10}}
\thicklines
% first item
\put(10,62){\line( 2,1){ 10}}
\put(10,62){\line( 0,-1){14}}
\put(10,62){\line(-2,1){ 10}}
% second item
\put(50,56){\line( 0,-1){8}}
\put(50,56){\line( 2,3){6}}
\put(50,56){\line(-2,3){6}}
\put(44,65){\line(1,0){12}}
\put(44,65){\line(-2,1){4}}
\put(56,65){\line( 2,1){4}}
\put(53,60.5){\vector( 2,3){  1}}
\put(47,60.5){\vector(2,-3){  1}}
\put(50,65){\vector(-1, 0){  1}}
\put(45,57){\makebox(0,0)[lb]{$R$}}
\put(55,57){\makebox(0,0)[rb]{$R$}}
\put(50,68){\makebox(0,0)[ct]{$R$}}
\thinlines
\put(47,68){\line(1,-1){6}}
\put(47,62){\line(1,1){6}}
\thicklines
% third item
\put(90,56){\line( 0,-1){8}}
\put(90,56){\line( 2,3){6}}
\put(90,56){\line(-2,3){6}}
%\put(84,65){\line(1,0){12}}
\put(84,65){\line(-2,1){4}}
\put(96,65){\line( 2,1){4}}
\put(93,60.5){\vector( 2,3){  1}}
\put(87,60.5){\vector(2,-3){  1}}
%\put(90,65){\vector(-1, 0){  1}}
\put(85,57){\makebox(0,0)[lb]{$R$}}
\put(95,57){\makebox(0,0)[rb]{$R$}}
%\put(90,68){\makebox(0,0)[ct]{$R$}}
\end{picture}
%}

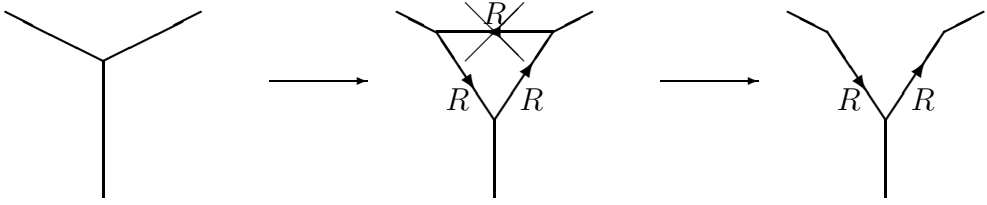
\centerline{{\bf Fig.~6.} ``Blowing up'' vertices of the initial graph.}

\hspace{10pt}

The matrices~$L$ and~$R$ then become the potential matrices
corresponding to the ``short'' oriented edges of the resulting graph~$\wtd T$. 

\begin{prop}\label{&prop1}~{\rm\cite{Fock}}
There is a one-to-one correspondence between the set of all
primitive closed paths $\{P_{z_1\cdots z_n}\}$ in the graph~$T$
and closed geodesics~$\{\gamma\}$ on the Riemann surface.
Moreover, the length~$l(\gamma)$ of a geodesic is determined by
the relation
\be
\label{&geod}
G(\gamma)\equiv 2\cosh \bigl(l(\gamma)/2\bigr)=\tr P_{z_1\cdots z_n}.
\ee
\end{prop}

\begin{remark}\label{&rem2}{\rm
Because both matrices (\ref{&Rz}) and (\ref{&Lz}) have the form
$\left\lbrace\begin{array}{cc} (+)&(-)\\ (-)&(+)\end{array}\right\rbrace$,
then {\it any product} of such matrices will be of the same form; meanwhile,
in a diagonal term of any such product, {\it all} summands enter with the
plus sign and, in an antidiagonal term, with the minus sign.
Then, for closed geodesics around holes (round--the--face geodesics), 
we obtain
$l(\gamma)=\left|\sum_{i\in I}^{}Z_i\right|$, where the sum ranges all 
boundary edges of the face (with the proper multiplicities).
}
\end{remark}

\begin{defin}\label{&def8}
{\rm The {\it Ihara--Selberg $L$-function} $L(u)$ for the fat
graph~$T$ is
\be
\label{&Lu}
L(u)=\prod_{\{\varpi\}}\det{}^{-1}(I-u^nP_{z_1\cdots z_n})
\ee
(cf. (\ref{&Lff}))
where the product runs over all primitive closed paths, $n$ being the
length of a path measured in terms of the distance on the universal
covering tree.}
\end{defin}

Then, from Theorem \ref{&th1}, we obtain that
the Ihara--Selberg $L$-function~{\rm(\ref{&Lu})} is a rational
function in variables $\e^{z_i/2}$ and $u$, and
\be
L(u)=\det{}^{-1}(I-T(u))=u^{r_0}\det{}^{-1}\Delta(u).
\label{&Tu}
\ee

The product (\ref{&Lu}) is absolutely convergent in the circle
$|u|<1/2 \min\{|\e^{-z_i}|\}$. However, following Theorem~\ref{&th1},
it possesses a unique analytic continuation into the whole~$\bf C$
except a finite number of singular points.

\subsection{Selberg trace formula and distribution of geodesics}

In the case of closed Riemann surface of constant negative curvature,
strong results concerning the distribution of closed geodesics over
lengths have been obtained (see~\cite{Poll,Series}).

\begin{defin}\label{&def9}
{\rm The {\it zeta-function} for closed geodesics on the (punctured)
surface~$S$ is
\be\label{&Z-function}
\zeta(s)=\prod\limits_{n}^{}\bigl(1-\e^{-shl_n}\bigr)^{-1},
\ee
where $h$ is a constant that are related to the asymptotic distribution
of geodesic lengths.
}
\end{defin}

If $\pi(T)$ is the number of closed
(primitive) geodesics with the length at most~$T$, then
the following assertion holds true (see, e.g.,~\cite{Poll,Series}).
\begin{claim}\label{&cl2}
There exists such constant~$h$ that
$$
\lim_{T\to\infty}\frac{\pi(T)}{\e^{hT}/hT}=1.
$$
\end{claim}

The proof of Claim~\ref{&cl2} follows from
the $\zeta$-function analyticity properties.
\begin{prop}\label{&prop2}~{\rm\cite{Poll}}
$\zeta(s)$ has an extension to $\Bbb C$ as a meromorphic function
such that\/{\rm:}

{\rm(a)} $\zeta(s)$ has no zeros or poles in $\Re(s)\ge1$, $s\ne1;$

{\rm (b)} $\zeta(s)$ has a simple pole at $s=1$.
\end{prop}

There are two alternative proofs of Proposition~\ref{&prop2}: one is based
on the celebrated Selberg trace formula, which, in its simplest form,
can be formulated as follows.

\begin{prop}{\rm (for proofs, see~\cite{Lax,Epstein}).}
Given the Laplace--Beltrami operator $\Delta$,
which acts on the hyperbolic
constant negative curvature metric space and is the linear second-order
\rom(unbounded\/\rom) partial differential operator with the discrete
spectrum $0=\lambda_0<\lambda_1\le\lambda_2\le\dots$, and the set of
closed geodesics with lengths $\{l_n\}$ on the Riemann surface, 
the Selberg trace formula relates these two sequences of real numbers
$\{\lambda_n\}$ and $\{l_n\}$ by the formula
$$
\sum_{n}^{}\hat f\Bigl(\sqrt{\lambda_n-1/4}\Bigr)+\int fdt=
\sum_{n}^{}c_n\bigl(f(l_n)\bigr)+2\frac{\hbox{\rm Area\,}(V)}{4\pi}
\int_{-\infty}^\infty r\tanh(r)\hat f(r)\,dr,
$$
where $f:{\Bbb R}\to{\Bbb R}$ is a $C^\infty$ function of compact support,
$\hat f$ is its Fourier transform, and 
$$
c_n\bigl(f(l_n)\bigr)=\sum_{k=1}^{\infty}\frac{l_nf(k l_n)}{\sinh(kl_n/2)},
$$
while $\hbox{\rm Area\,}(V)/2\pi=2(g-1)$ for a compact surface.
\end{prop}

Alternative proof presented in~\cite{Poll} just uses the methods of symbolic
dynamics and can be rather close to the combinatorial graph description.
Here we formulate two problems.

\begin{prob}
{\rm To relate the graph description of zeta- (or $L$-)function (\ref{&Lff})
of the Teichm\"uller space with the standard zeta-function (\ref{&Z-function}).
}
\end{prob}

\begin{prob}
{\rm
Few explicit formulas containing the geodesics in the open Riemann 
surface case are known. Worth mentioning is the paper~\cite{McShane},
where the formulas concerning sets of simple (\ie non-self-intersecting)
geodesics were obtained for a punctured torus and for a pair of pants.
Are there generalizations of these formulas to a case of surfaces with 
holes?
}
\end{prob}

\subsection{Classical projective (modular) transformations}

In~\cite{Fock}, the projective transformations on graphs that
are the mapping class group (modular) transformations
were obtained. They correspond to natural operations called the
{\it flips}, or {\it Whitehead moves}, 
which are elementary transitions between
graphs (actually, between neighbor cells of the simplicial complex
whose higher dimensional cells label combinatorial types of 
three-valent fat graphs
of the given Riemannian genus
$g$ and~$n$). The corresponding
transformation of the variables $Z_\alpha$ is nonlinear 
(see~\cite{Fock,Ch-Fock} for the explicit expressions), but
the {\it geodesic length} is a {\it modular invariant} function.

\begin{lemma}\label{&lem22}
At $u=1$, the function $L(u)$ is modular invariant,
i.e., does not depend on the particular form of the representing graph.
\end{lemma}

Note, however, that
the graph length~$n$ of a geodesic varies under the modular transformations,
so we are still unable to define a complete analogue of the zeta function
(\ref{&Z-function}) in terms of graphs. 
However, constructing $L$-functions (\ref{&Lu})
makes sense in the symbolic
dynamics setting where one customarily introduce additional 
modular-noninvariant partition of 
a Riemann surface, which would corresponds to a graph decomposition
(see Chaps.~5 and~6 in~\cite{Series}).

\begin{remark}\label{&rem3}{\rm
As for the function $L^{-1}(u)$, 
the determinant in (\ref{&Lu}) is a Laurent polynomial of no more than
second order in $\e^{\pm z_i/2}$ for each $z_i$. Moreover, from 
Lemma~\ref{&lem2} it follows that the modular-invariant expression for $L(1)$ 
can depend only on the modular invariants ``perimeters'' of holes.
This imposes severe restrictions 
on a possible form of $L^{-1}(1)$. Say, for the moduli spaces $M_{g,1}$ of 
the Riemann surfaces with one puncture (hole), 
there exists only one such parameter,
$p_1$. Taking into account the global symmetry $x_i\to -x_i$,
we find that
$$
L^{-1}_{g,1}(1)=a\cosh(p_1)+b\cosh(p_1/2)+c,\quad a+b+c=0,
$$
where only the coefficients $a$,
$b$, and~$c$ depend on the genus of the surface.}
\end{remark}

\begin{example}\label{&exam2}{\rm
1. For the moduli space $M_{1,1}$ (the torus with one puncture),
there are three parameters $x$, $y$, and~$z$ and one modular invariant
$p_1=x+y+z$,
$$
L^{-1}_{1,1}(1)=-8\bigl(\cosh(x+y+z)-1\bigr).
$$

2. The moduli space $M_{0,3}$ (the sphere with three punctures).
There are three parameters $x$, $y$, and~$z$, and three restrictions
$x+y=p_3$, $x+z=p_2$, $y+z=p_1$. Then}
$$
L^{-1}_{0,3}(1)=-2^5\left(\cosh\left(\frac{x+y}{2}\right)-1\right)
\left(\cosh\left(\frac{x+z}{2}\right)-1\right)
\left(\cosh\left(\frac{y+z}{2}\right)-1\right).
$$
\end{example}

\newsection{Conclusions}
Our main result is the relation (\ref{&*11}) connecting the
$L$-functions with the scattering data. It provides an analogue of 
the Selberg trace formula for the discrete non-compact case of 
the graphs.  

Spectral properties of the $S$-matrix studied in Sec.~4 deserve
further investigation, especially for the case of unitary
and Schr\"odinger type potentials.

One can hope to apply the technique of this paper to the continuous 
non-compact case in order to obtain analogous formulas for the continuum 
scattering processes.

Eventually, some $r$-matrix structure must be hidden in these scattering 
processes. It is interesting how it can be interpreted from the 
standpoint of the integrable models.

After this paper has been completed we were aware of paper~\cite{Akk}, where
the similar results concerning magnetic fluxes on graphs have been obtained
within the functional integral standpoint.

\newsection{Acknowledgements}
I am grateful to S.~P.~Novikov and the participants of his seminar
at Steklov Mathematical Institute for the useful discussion and
for the proposal to write this paper.

This paper was supported by the Russian Foundation for Basic Research
(Grant No.~98-01-00327).

\end{document}